\documentclass[11pt,a4paper]{article}

\usepackage[T1]{fontenc}
\usepackage{amsmath, amssymb, amsfonts}
\usepackage{graphicx}
\usepackage{geometry}
\usepackage{cite}
\usepackage{xcolor}
\usepackage[colorlinks=true, citecolor=blue, linkcolor=blue, urlcolor=blue]{hyperref}
\usepackage{float}
\usepackage{booktabs}
\usepackage[section]{placeins}
\usepackage{authblk}
\usepackage{subcaption}
\usepackage[nopatch=footnote]{microtype}
\usepackage{titlesec}

\title{\textbf{Parameter Scan of Multi-Fluid Equilibria in Rotating \texorpdfstring{$p\text{-}^{11}\text{B}$}{p-11B} Plasmas: Effects on Fusion Power and Bremsstrahlung Losses}}
\author[1,2]{Xingyu Li}
\author[2,3]{Huasheng Xie\textsuperscript{\textdagger}}
\author[1]{Lai Wei}
\author[1]{Zheng-Xiong Wang\textsuperscript{\textdagger}}

\affil[1]{\small Key Laboratory of Materials Modification by Laser, Ion, and Electron Beams of the Ministry of Education, School of Physics, Dalian University of Technology, Dalian 116024, China}
\affil[2]{\small ENN Science and Technology Development Co., Ltd., Langfang 065001, China}
\affil[3]{\small Beijing VeloAlpha Technology Co., Ltd., Beijing, 100080, China}
\date{\today}

\begin{document}

\maketitle

{
  \renewcommand{\thefootnote}{\textdagger}
  \footnotetext{Corresponding authors. E-mail: huashengxie@gmail.com (H. Xie), zxwang@dlut.edu.cn (Z.-X. Wang).}
}

% ====== Abstract ======
\begin{abstract}
We present VEQ-MF, a fast spectral parameter-scan framework for two-dimensional axisymmetric multi-fluid equilibria with prescribed species-dependent toroidal rotation. The solver couples generalized Boltzmann density responses, quasineutral electrostatic polarization, and a generalized Grad--Shafranov equation, extending reduced-parameter Grad--Shafranov and VEQ formulations to multi-species rotating equilibria. Rotating $p\text{-}^{11}\text{B}$ spherical-tokamak configurations are used as a demanding test case. Independent scans of the proton and boron rotation frequencies are performed in EHL-2 and EHL-3B geometries. The computed fields are then post-processed to obtain fusion power from a drift-Maxwellian reaction-rate coefficient and bremsstrahlung power from an analytical radiation model. Three in-range EHL-3B finite-difference benchmarks give global stored-energy, bremsstrahlung-power, and fusion-power differences of $1.7$--$3.4\%$, while a representative convergence check shows sub-percent sensitivity to increasing the spectral-parameter number and negligible sensitivity to Gaussian-grid refinement. The core equilibrium solve remains fast for repeated scans, with representative nonzero EHL-3B cases requiring $0.032$--$0.050$~s per point in MATLAB, excluding post-processing, interpolation, plotting, and file export.

The scans identify two competing multi-fluid effects. Under iso-rotation, outward boron accumulation increases the volume-integrated $n_e^2$, so the fusion-to-bremsstrahlung power ratio $\mathcal{R}_{\mathrm{fb}}$ decreases with increasing rotation. Species-dependent toroidal rotation weakens centrifugal polarization and lowers bremsstrahlung power, while the relative toroidal flow in the larger EHL-3B geometry raises the drift-Maxwellian reaction-rate coefficient and thereby modifies fusion power. A diagnostic decomposition shows that the proton--boron overlap integral decreases monotonically with increasing differential rotation, whereas the density-overlap-weighted effective reaction-rate coefficient rises steadily. Within the prescribed-profile scan, these combined effects bring the bremsstrahlung-only power ratio to an $\mathcal{R}_{\mathrm{fb}}=1$ crossing; in a fixed-$\Omega_{p0}$ EHL-3B scan, $\mathcal{R}_{\mathrm{fb}}$ rises from 0.18 at iso-rotation to about 1.3 at the largest imposed relative toroidal flow. The results provide a benchmarked equilibrium-scan workflow for separating the density-overlap and reaction-rate-coefficient contributions to fusion power in rotating multi-fluid plasmas.
\end{abstract}

\vspace{0.5cm}

% ====== Main Text ======
\section{Introduction}
\label{sec:introduction}

The $p\text{-}^{11}\text{B}$ reaction is neutron-lean, but its cross section peaks near $6.0\times10^2$~keV.
For $p\text{-}^{11}\text{B}$ plasmas, appreciable Maxwellian reaction-rate coefficients therefore require ion temperatures of several hundred keV~\cite{putvinski2019, tentori2023}.
At such temperatures, electrons are heated as well, and relativistic bremsstrahlung becomes a severe loss channel.
This is one reason conventional thermal $p\text{-}^{11}\text{B}$ gain is difficult~\cite{rider1995, dawson1981}. Rider also emphasized the power cost of maintaining non-equilibrium plasma states; the momentum-exchange estimate in Appendix~\ref{app:momentum_exchange} is used below to place the prescribed differential-rotation states in that context.
These constraints motivate model studies in which the relative energy of reacting ions is increased without raising $T_e$ by the same amount.
Alpha-particle channeling is a separate route for recovering fusion-product energy before it thermalizes with electrons~\cite{fisch1992, mlodik2022}.

Neutral beam injection in spherical tokamaks (STs) can drive toroidal rotation while supporting plasma performance, as reported in NSTX experiments~\cite{kaye2005progress}. The $\sim 1.0\times10^6$~rad/s rotation amplitudes used below are theoretical scan parameters, not direct reproductions of NSTX rotation amplitudes.
For plasmas with large mass disparity, such as $p\text{-}^{11}\text{B}$ ($m_B \approx 11 m_p$), the single-fluid approximation becomes incomplete once the rotation speed approaches the sound speed.
Fulop and Helander analyzed centrifugal redistribution in rotating impurity plasmas~\cite{fulop1999nonlinear}.
Casson \textit{et al.} included centrifugal effects in gyrokinetic calculations~\cite{casson2010gyrokinetic}.
In the present setting, the same mass-dependent centrifugal force drives heavy ions toward the low-field side (LFS).
The resulting charge imbalance produces an electrostatic potential $\Phi$ that couples the electron density to the ion distribution in the VEQ-MF model~\cite{xie2026development}.
Because bremsstrahlung scales with $n_e^2$, this redistribution can increase radiative losses relative to zero-dimensional or single-fluid estimates.

If the rotation frequencies of protons and boron differ, two effects appear. First, reducing $\Omega_B$ relative to $\Omega_p$ weakens the centrifugal force on boron, reduces the poloidal density asymmetry, and lowers the volume-integrated bremsstrahlung. We describe this behavior below as a reduction of centrifugal polarization. Second, the relative toroidal flow speed $\Delta u = R |\Omega_p - \Omega_B|$ corresponds to a directed drift kinetic energy $E_{\mathrm{d}} = m_{\mathrm{r}} (\Delta u)^2/2$ in the center-of-mass frame. Under a drift-Maxwellian ansatz for the relative velocity distribution~\cite{xie2024upper}, this shift increases the reaction-rate coefficient $\langle\sigma v\rangle_{\mathrm{DM}}$ without increasing $T_e$ by the same amount. Related kinetic effects in idealized configurations were analyzed by Kolmes \textit{et al.}~\cite{kolmes2021fusion}. The present work brings centrifugal redistribution and the resulting drift-induced fusion-power contribution into the same two-dimensional multi-fluid equilibrium scan, and then decomposes their effects on $\mathcal{R}_{\mathrm{fb}}$.

The purpose of this work is not to assess reactor-level feasibility for the $p\text{-}^{11}\text{B}$ route. Instead, $p\text{-}^{11}\text{B}$ is used as a demanding model case for testing a parameter-scan multi-fluid equilibrium calculation under prescribed profiles. This distinction is intentional. The paper presents a matched-input parameter scan: VEQ-MF generates two-dimensional multi-fluid equilibria, a finite-difference solver provides numerical cross-validation, and the computed fields are used to diagnose how density redistribution and the drift-Maxwellian reaction-rate coefficient combine in the fusion-power integral. This diagnostic decomposition is the central objective of the analysis. VEQ-MF builds on reduced-parameter GS representations~\cite{xie2026minimumgs} and the VEQ parametric spectral solver~\cite{zhang2026veq}, and extends them to the coupled system of generalized Boltzmann distributions and the Grad--Shafranov equation. The solver is then used to scan the $(\Omega_p,\Omega_B)$ plane for two ST geometries, EHL-2 ($R_0 = 1.1$~m) and EHL-3B ($R_0 = 3.2$~m)~\cite{liu2024enn}. For each case, fusion power is evaluated with a drift-Maxwellian reaction-rate coefficient, relativistic bremsstrahlung power is calculated, and the bremsstrahlung-only power ratio $\mathcal{R}_{\mathrm{fb}}$ is mapped.

The species rotation profiles and temperature profiles are prescribed as control parameters; inter-species momentum exchange, the power required to sustain relative toroidal flow, and frictional thermalization are not modeled. The calculations address the equilibrium consequences of imposed differential rotation rather than the transport-limited sustainment of such states. The reported $\mathcal{R}_{\mathrm{fb}}$ values therefore characterize this prescribed-profile equilibrium model.

Section~\ref{sec:model} describes the multi-fluid equilibrium model, the VEQ-MF solver, and the power-evaluation method. Section~\ref{sec:spatial} examines the spatial structure of the equilibria. Section~\ref{sec:parameter} presents the dependence of $P_{\mathrm{fus}}$, $P_{\mathrm{brem}}$, and $\mathcal{R}_{\mathrm{fb}}$ on the rotation parameters. Section~\ref{sec:mechanisms} analyzes the corresponding physical mechanisms. Conclusions are given in Section~\ref{sec:conclusion}. The appendices discuss the prescribed profiles, the high-rotation equilibrium limit, the momentum-exchange power scale, numerical implementation, and supplementary modal sensitivity.

\section{Multi-Fluid Equilibrium Model}
\label{sec:model}

The multi-fluid equilibrium framework of Xie \textit{et al.}~\cite{xie2026development} is used to calculate centrifugal species-density redistribution and electrostatic polarization in rotating $p\text{-}^{11}\text{B}$ plasmas. The model retains the centrifugal force associated with toroidal rotation and the self-consistent polarization field, while neglecting poloidal flow inertia to avoid transonic numerical singularities~\cite{casson2010gyrokinetic}. More general treatments of rotating plasma equilibria can be found in Refs.~\cite{maschke1980exact, hameiri1983equilibrium, guazzotto2004numerical}.

\subsection{Governing Equations}

We consider an axisymmetric multi-species plasma consisting of protons (p), boron ions ($^{11}\text{B}$), and electrons (e). The steady state of each species $s$ is described by the multi-fluid momentum equation
\begin{equation}
    m_s n_s (\vec{u}_s \cdot \nabla) \vec{u}_s = -\nabla p_s + n_s q_s (\vec{E} + \vec{u}_s \times \vec{B}),
    \label{eq:momentum}
\end{equation}
where $m_s$, $n_s$, $\vec{u}_s$, $p_s$, and $q_s$ denote the mass, density, flow velocity, scalar pressure, and charge of species $s$, respectively. The electrostatic field is $\vec{E} = -\nabla \Phi$, and the axisymmetric magnetic field is written as $\vec{B} = \nabla\phi \times \nabla\psi + F(\psi)\nabla\phi$.

Three assumptions reduce the system to a tractable form:
\begin{enumerate}
    \item \textbf{Dominant toroidal rotation}: poloidal flow inertia is neglected. The flow is taken to be purely toroidal, $\vec{u}_s = R \Omega_s(\psi)\hat{\phi}$, with prescribed flux-surface angular frequency $\Omega_s(\psi)$.
    \item \textbf{Flux-surface isothermality}: high parallel thermal conductivity gives $T_s = T_s(\psi)$, and the scalar pressure is $p_s = n_s k_B T_s$.
    \item \textbf{Massless electron limit}: electron inertial centrifugal effects are neglected, so $W_e^{\mathrm{cf}} \to 0$.
\end{enumerate}

These assumptions define a constrained-flow equilibrium model. In the highest-rotation EHL-3B cases, the boron Mach number reaches the transonic or supersonic range, where a fully self-consistent rotating equilibrium may require poloidal-flow inertia and transport-level closure. The high-rotation tail reported below is therefore interpreted as the response of the prescribed pure-toroidal-flow model, not as a complete supersonic rotating-equilibrium prediction or a physical stability boundary.

Projecting Eq.~\eqref{eq:momentum} along $\vec{B}$ removes the Lorentz force. Integration along a field line then gives the generalized Boltzmann distribution for the species density,
\begin{equation}
    n_s(R, \psi) = N_s(\psi) \exp \left[ \frac{W_s^{\mathrm{cf}}(R, \psi) - q_s \Phi(R, \psi)}{k_B T_s(\psi)} \right],
    \label{eq:density_dist}
\end{equation}
where $N_s(\psi)$ is the reference density at the magnetic axis $R_0$, and $\Phi(R,\psi)$ is the polarization potential. The centrifugal potential is
\begin{equation}
    W_s^{\mathrm{cf}}(R, \psi) = \frac{1}{2} m_s \Omega_s^2(\psi) (R^2 - R_0^2).
    \label{eq:centrifugal_potential}
\end{equation}

Because $m_B \approx 11 m_p$, the centrifugal potential of $^{11}\text{B}$ is proportionally larger than that of protons at the same $\Omega_s$. This favors boron accumulation at larger $R$. The resulting electrostatic polarization generates $\Phi$, which is constrained by quasi-neutrality,
\begin{equation}
    \sum_s q_s n_s(R,\psi,\Phi) = 0.
    \label{eq:quasineutrality}
\end{equation}

\subsection{Generalized Grad--Shafranov Equation and Nonlinear Coupling}

Mechanical equilibrium across flux surfaces is described by the generalized Grad--Shafranov (GS) equation,
\begin{equation}
    \Delta^* \psi = -\mu_0 R^2 \left( \frac{\partial P_{\mathrm{tot}}}{\partial \psi} \right)_R - F \frac{dF}{d\psi},
    \label{eq:generalized_gs}
\end{equation}
where the total pressure
\[
P_{\mathrm{tot}}(R,\psi)=\sum_s n_s(R,\psi) k_B T_s(\psi)
\]
is a two-dimensional quantity. The source term $(\partial P_{\mathrm{tot}}/\partial \psi)_R$ must therefore be expanded with the chain rule. Following Ref.~\cite{xie2026development}, we define an auxiliary variable $A_s$ that includes toroidal flow shear through $\Omega_s \Omega_s'$,
\begin{equation}
    A_s = \frac{N_s'}{N_s}
    - \frac{T_s'}{T_s}\left(\frac{W_s^{\mathrm{cf}} - q_s \Phi}{k_B T_s}\right)
    + \frac{m_s \Omega_s \Omega_s' (R^2 - R_0^2)}{k_B T_s}.
    \label{eq:auxiliary_As}
\end{equation}

Differentiating the quasi-neutrality condition, Eq.~\eqref{eq:quasineutrality}, with respect to $\psi$ at fixed $R$ gives the closure relation for the polarization-potential gradient,
\begin{equation}
    \left( \frac{\partial \Phi}{\partial \psi} \right)_R
    =
    \frac{\sum_s q_s n_s A_s}
    {\sum_s q_s^2 n_s /(k_B T_s)}.
    \label{eq:phi_gradient}
\end{equation}

The corresponding species-density gradient and total-pressure gradient are
\begin{equation}
    \left(\frac{\partial n_s}{\partial \psi}\right)_R
    = n_s \left[
    A_s - \frac{q_s}{k_B T_s} \left( \frac{\partial \Phi}{\partial \psi} \right)_R
    \right],
    \label{eq:density_gradient}
\end{equation}
and
\begin{equation}
    \left( \frac{\partial P_{\mathrm{tot}}}{\partial \psi} \right)_R
    = \sum_s \left[
    k_B T_s \left( \frac{\partial n_s}{\partial \psi} \right)_R
    + n_s k_B T_s'
    \right].
    \label{eq:pressure_gradient}
\end{equation}

Equations~\eqref{eq:auxiliary_As}--\eqref{eq:pressure_gradient}, together with Eq.~\eqref{eq:quasineutrality}, define the nonlinear feedback in the model. Centrifugal forcing generates polarization, polarization modifies the pressure gradient through $\partial \Phi/\partial \psi$ and $\Omega_s'$, and the modified pressure gradient in turn reshapes $\psi$ and the spatial distributions.

\subsection{VEQ-MF Spectral Solver}

The coupled system is solved numerically with VEQ-MF, a multi-species extension of the reduced-parameter GS representation and VEQ spectral-solver framework~\cite{xie2026minimumgs,zhang2026veq}. The solver treats the stiff exponential density response at high rotation through a two-layer nonlinear structure with continuation (homotopy):
\begin{enumerate}
    \item \textbf{Local quasi-neutrality (inner loop):} Eq.~\eqref{eq:quasineutrality} is solved pointwise with a vectorized Newton--Raphson method using an analytic Jacobian.
    \item \textbf{Global GS operator (outer loop):} the generalized GS system is solved globally with shifted Chebyshev polynomials. A hybrid Broyden method is used for the 16 spectral parameters. When the Broyden update fails, the solver falls back to a finite-difference Jacobian regularized with a singular-value-decomposition (SVD) pseudo-inverse.
    \item \textbf{Multi-stage homotopy continuation}: the rotation frequencies $\Omega_s$ are increased incrementally, using converged low-rotation states as initial guesses for the next step.
\end{enumerate}

Details of the spectral discretization and nonlinear iteration are given in Appendix~\ref{app:numerical_solver}.

The implementation is designed for repeated equilibrium solves in parameter scans. Hereafter, the ratio of the prescribed boron and proton core rotation frequencies is denoted by $\gamma_{Bp}\equiv\Omega_{B0}/\Omega_{p0}$. In a representative MATLAB R2024b timing test on the 16$\times$16 Gauss grid used here, the zero-rotation origin point required $0.0020$~s. Three nonzero EHL-3B points from the absolute-frequency $(\Omega_{p0},\Omega_{B0})$ scan, sampled along $\gamma_{Bp}=0.20$, converged with per-point equilibrium-solve times of $0.032$--$0.050$~s, with a median of $0.041$~s. The initial homotopy continuation and 25~MA current-matching steps required $0.15$~s and $0.066$~s, respectively. These timings include the nonlinear equilibrium solve and the $I_p$ feedback calls to the multi-fluid kernel, but exclude power post-processing, interpolation, plotting, and file export. They document the cost of the equilibrium-solve stage on the tested workstation and are not presented as an independent performance assessment.

\subsection{Numerical Benchmark}

The VEQ-MF solutions are compared with a high-resolution finite-difference (FDM) solver under matched input parameters, including the same device geometry, density amplitudes, temperature amplitudes, rotation frequencies, and 25~MA current target.

\subsubsection*{Single-point high-rotation stress test}

\begin{figure}[!htbp]
    \centering
    \includegraphics[width=0.9\textwidth]{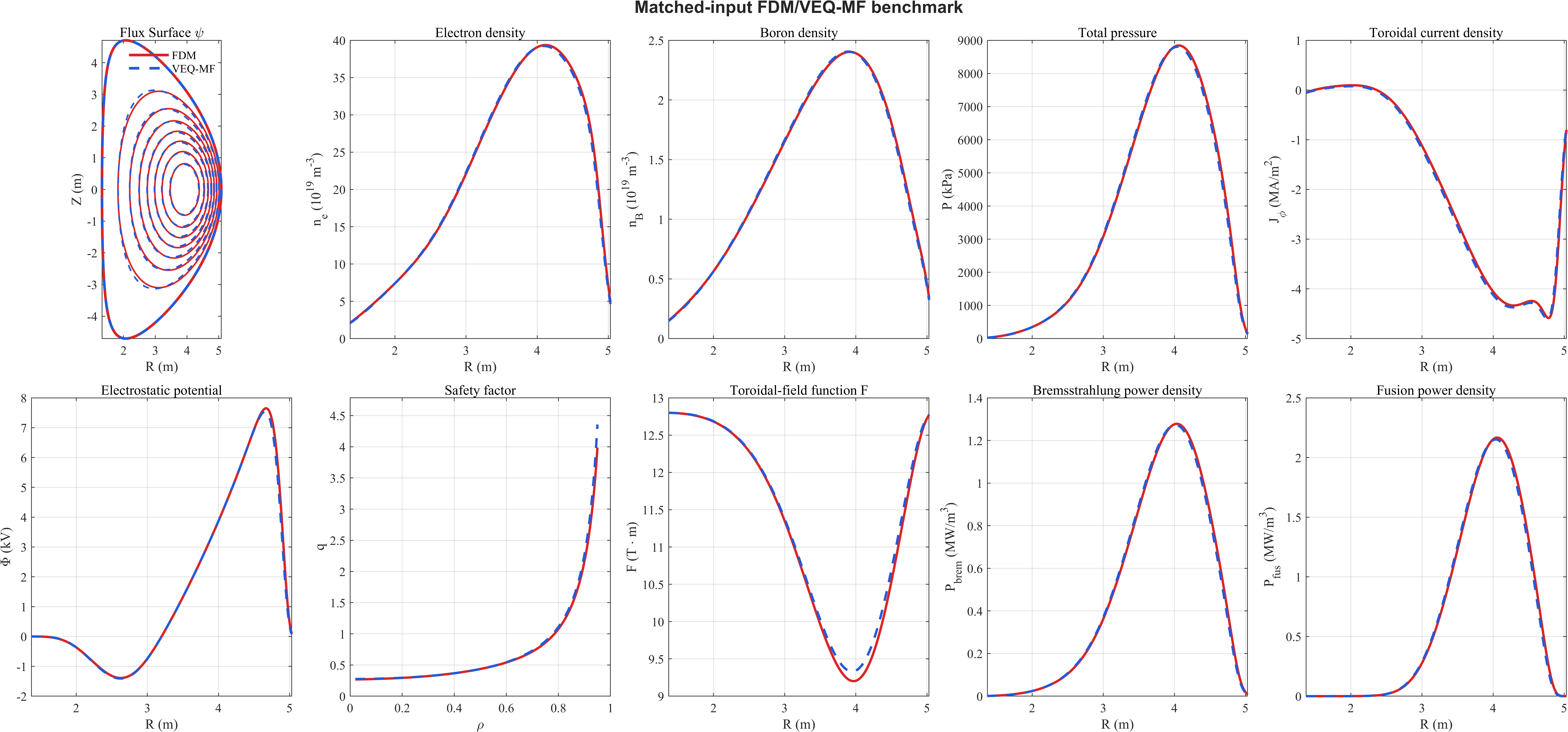}
    \caption{Matched-input FDM/VEQ-MF stress-test benchmark for the high-rotation EHL-3B case with $\Omega_p = 1.0 \times 10^6$~rad/s and $\Omega_B = 0.20 \times 10^6$~rad/s. The global stored energy, bremsstrahlung power, and fusion power agree within 2.9\%, 2.8\%, and 3.7\%, respectively.}
    \label{fig:benchmark_profiles}
\end{figure}

\begin{figure}[!htbp]
    \centering
    \includegraphics[width=0.95\textwidth]{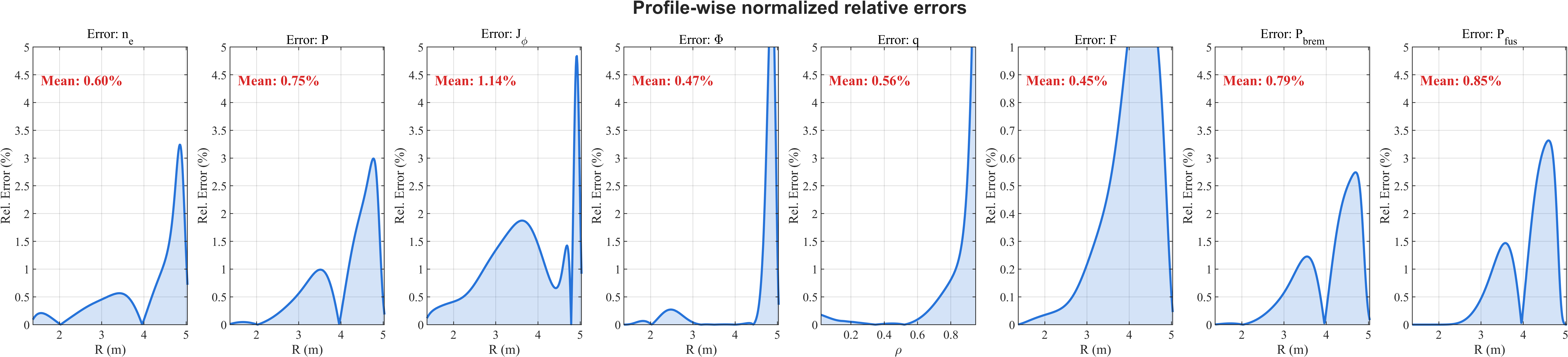}
    \caption{Profile-wise relative errors in the matched-input benchmark, computed with respect to the FDM reference and normalized by the maximum amplitude of each corresponding FDM profile.}
    \label{fig:benchmark_errors}
\end{figure}

Figures~\ref{fig:benchmark_profiles} and \ref{fig:benchmark_errors} show the original high-rotation EHL-3B stress-test case with $\Omega_p = 1.0 \times 10^6$~rad/s and $\Omega_B = 0.20 \times 10^6$~rad/s. Because this point is slightly above the reported EHL-3B scan boundary of $0.90$~Mrad/s, it is used here as a nonlinear stress test rather than as the sole validation point for the plotted scan space.

For each profile, the relative error is computed with respect to the FDM reference solution and normalized by the maximum amplitude of the corresponding reference profile. The comparison provides numerical cross-validation for this high-rotation stress-test case.

The stress-test global integral comparison gives $W_{\mathrm{mhd}}\simeq1.8\times10^3$~MJ for FDM and $1.7\times10^3$~MJ for VEQ-MF, a difference of $2.9\%$. The corresponding values are $P_{\mathrm{brem}}\simeq1.1\times10^2$~MW and $1.0\times10^2$~MW, differing by $2.8\%$, and $P_{\mathrm{fus}}\simeq1.2\times10^2$~MW for both solvers, differing by $3.7\%$. A supplementary modal-sensitivity check at the same stress-test input is given in Appendix~\ref{app:modal_sensitivity}.

\subsubsection*{Three-point in-range cross-validation}

To test the reported scan range directly, three additional FDM/VEQ-MF comparisons were run on the $\gamma_{Bp}=0.20$ EHL-3B line at $\Omega_{p0}=0.30$, $0.60$, and $0.90$~Mrad/s. Table~\ref{tab:in_range_benchmark} shows that the global stored-energy, bremsstrahlung-power, fusion-power, and bremsstrahlung-only $\mathcal{R}_{\mathrm{fb}}$ differences remain at the few-percent level across these in-range points. On the common midplane interval, the largest normalized profile error among $n_e$, $n_B$, $\Phi$, $P_{\mathrm{brem}}$, and $P_{\mathrm{fus}}$ stays below $6.3\%$.

\begin{table}[!htbp]
    \centering
    \caption{Matched-input in-range FDM/VEQ-MF benchmark along the EHL-3B $\gamma_{Bp}=0.20$ line. The power ratio is the bremsstrahlung-only post-processing quantity $\mathcal{R}_{\mathrm{fb}}=P_{\mathrm{fus}}/P_{\mathrm{brem}}$.}
    \label{tab:in_range_benchmark}
    \resizebox{\textwidth}{!}{%
    \begin{tabular}{cccccccc}
        \toprule
        $\Omega_p$ & $\Omega_B$ & $\Delta W$ & $\Delta P_{\mathrm{brem}}$ & $\Delta P_{\mathrm{fus}}$ & $\mathcal{R}_{\mathrm{fb}}$ (FDM) & $\mathcal{R}_{\mathrm{fb}}$ (VEQ-MF) & $\Delta\mathcal{R}_{\mathrm{fb}}$ \\
        (Mrad/s) & (Mrad/s) & & & & & & \\
        \midrule
        0.30 & 0.060 & 1.7\% & 1.9\% & 2.4\% & 0.71 & 0.70 & 0.54\% \\
        0.60 & 0.12 & 2.0\% & 2.1\% & 2.8\% & 0.83 & 0.82 & 0.69\% \\
        0.90 & 0.18 & 2.6\% & 2.6\% & 3.4\% & 1.1 & 1.0 & 0.86\% \\
        \bottomrule
    \end{tabular}
    }
\end{table}

\begin{figure}[!htbp]
    \centering
    \includegraphics[width=0.9\textwidth]{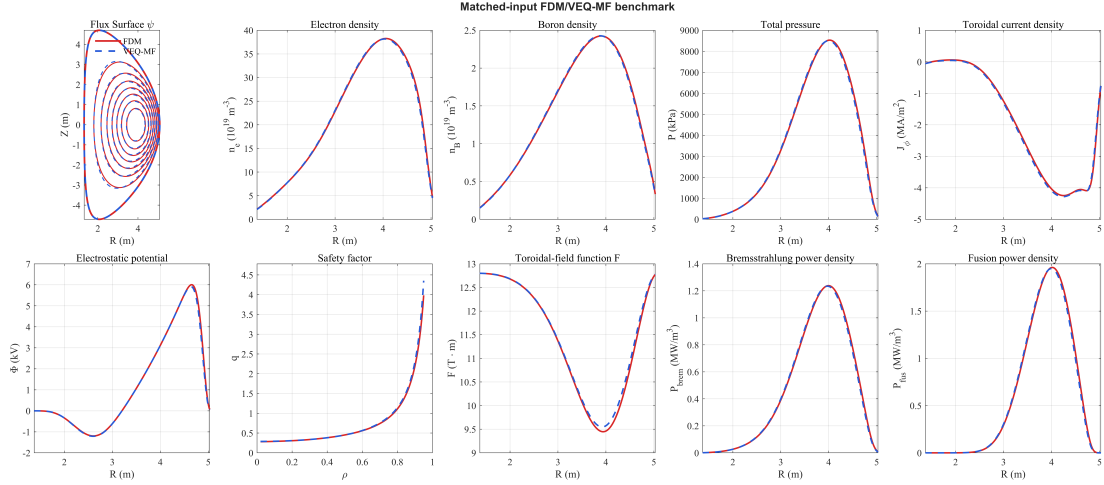}
    \caption{Matched-input FDM/VEQ-MF midplane comparison for the in-range EHL-3B benchmark point with $\Omega_{p0}=0.90$~Mrad/s, $\Omega_{B0}=0.18$~Mrad/s, and $\gamma_{Bp}=0.20$.}
    \label{fig:inrange_benchmark_op09}
\end{figure}

\begin{figure}[!htbp]
    \centering
    \includegraphics[width=0.95\textwidth]{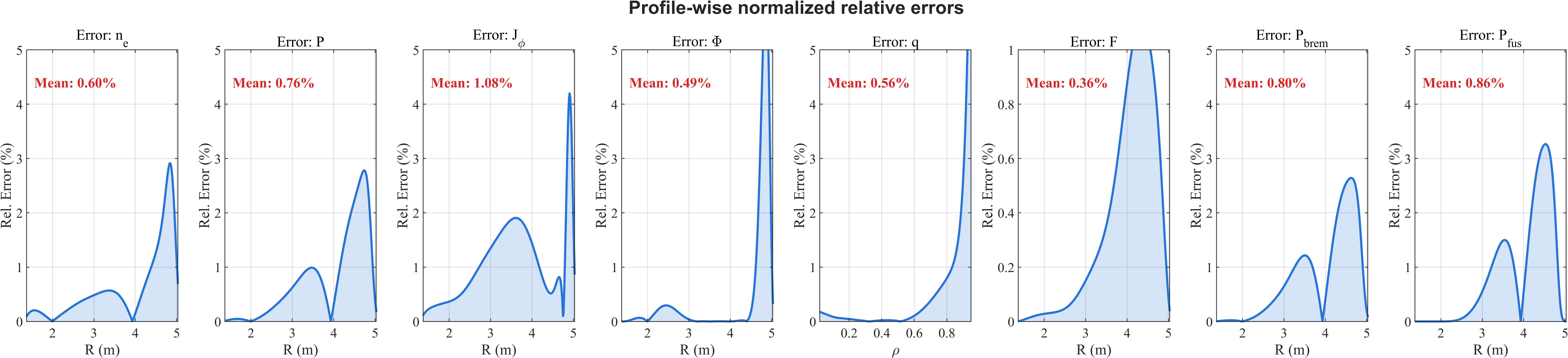}
    \caption{Profile-wise relative errors for the in-range EHL-3B benchmark point with $\Omega_p=0.90$~Mrad/s and $\Omega_B=0.18$~Mrad/s. Errors are computed with respect to the FDM reference solution and normalized by the maximum FDM reference amplitude of each profile.}
    \label{fig:inrange_benchmark_op09_errors}
\end{figure}

The scan-boundary point $\Omega_p=0.90$~Mrad/s and $\Omega_B=0.18$~Mrad/s is the most demanding of the three in-range benchmark cases. Figures~\ref{fig:inrange_benchmark_op09} and \ref{fig:inrange_benchmark_op09_errors} show the corresponding midplane profiles and normalized profile errors. These figures complement the $1.0/0.20$~Mrad/s stress-test benchmark in Figs.~\ref{fig:benchmark_profiles} and \ref{fig:benchmark_errors} by showing the solver comparison directly at the upper end of the reported scan range.

\subsubsection*{In-range convergence check}

A representative VEQ-MF convergence check was also carried out at the in-range high-rotation point $\Omega_p=0.90$~Mrad/s and $\Omega_B=0.18$~Mrad/s. The standard order-3 Chebyshev representation with 16 spectral parameters was compared with an order-5 representation with 24 spectral parameters, and the 16$\times$16 Gaussian quadrature grid was compared with a 32$\times$32 grid. Table~\ref{tab:convergence_in_range} shows that the reported global quantities vary by less than $0.12\%$ under the modal refinement and less than $0.0020\%$ under the grid refinement for this representative point.

\begin{table}[!htbp]
    \centering
    \caption{Representative in-range VEQ-MF modal and Gaussian-grid convergence check for EHL-3B with $\Omega_p=0.90$~Mrad/s and $\Omega_B=0.18$~Mrad/s. Absolute values are rounded for scale; relative differences are computed from the unrounded runs.}
    \label{tab:convergence_in_range}
    \resizebox{\textwidth}{!}{%
    \begin{tabular}{lcccc}
        \toprule
        \textbf{Run} & $W_{\mathrm{mhd}}$ (MJ) & $P_{\mathrm{brem}}$ (MW) & $P_{\mathrm{fus}}$ (MW) & $\mathcal{R}_{\mathrm{fb}}$ \\
        \midrule
        Baseline (order 3, 16 parameters, $16\times16$) & $1.6\times10^3$ & $1.0\times10^2$ & $1.1\times10^2$ & $1.0$ \\
        Modal refinement (order 5, 24 parameters) & $1.6\times10^3$ & $1.0\times10^2$ & $1.1\times10^2$ & $1.0$ \\
        Rel.\ diff.\ (modal) & 0.041\% & 0.067\% & 0.11\% & 0.046\% \\
        Grid refinement ($32\times32$) & $1.6\times10^3$ & $1.0\times10^2$ & $1.1\times10^2$ & $1.0$ \\
        Rel.\ diff.\ (grid) & 0.0019\% & 0.0011\% & 0.00080\% & 0.00030\% \\
        \bottomrule
    \end{tabular}
    }
\end{table}

Together, the single-point stress test, the three-point in-range cross-validation, the in-range convergence check, and the supplementary modal-sensitivity check indicate that the global quantities used in the subsequent parameter scans are numerically stable at the level required for the reported trends. These benchmarks are used for numerical cross-validation only. They support the few-percent agreement of the two discretizations and the stability of the VEQ-MF global quantities at the representative high-rotation point, but they are not an exhaustive validation of every local profile feature or every point in the two-dimensional scan space. The formal parameter scans reported below remain restricted to the numerically accessible range discussed in Appendix~\ref{app:rotation_limit}.

\subsection{Power-Ratio Evaluation}

The converged two-dimensional fields $(n_s(R,Z), \Phi(R,Z), \psi(R,Z))$ are used to evaluate volume integrals with $dV = 2\pi R\, dR\, dZ$. The quantities of interest are bremsstrahlung loss and fusion power. These evaluations are carried out on equilibria obtained with prescribed species rotation profiles. In the main scan figures,
\[
\mathcal{R}_{\mathrm{fb}}=\frac{P_{\mathrm{fus}}}{P_{\mathrm{brem}}}
\]
is therefore a bremsstrahlung-only power ratio. Other radiative and transport losses, auxiliary power, and the frictional power required to maintain finite relative toroidal flow between ion species are not included in this denominator. The resulting $\mathcal{R}_{\mathrm{fb}}$ values are reduced power-balance indicators, not reactor-level net-gain estimates. Appendix~\ref{app:momentum_exchange} gives an order-of-magnitude estimate of the collisional momentum-exchange power scale for the largest imposed relative-toroidal-flow EHL-3B case.

We also note that the prescribed temperature profiles $T_s(\psi)$ are used here as part of a reduced equilibrium closure. For finite relative toroidal flow between ion species, Coulomb collisions can convert directed kinetic energy into thermal energy and thereby modify the ion-temperature profiles. Such thermalization and the associated temperature evolution are not treated self-consistently in the present model. A fully self-consistent treatment would require coupling the equilibrium calculation to transport and energy-balance equations, which is beyond the scope of this work. The present results therefore isolate the effects of centrifugal redistribution and the adopted drift-Maxwellian reaction-rate coefficient on the post-processed powers under prescribed temperature profiles.

Unless otherwise stated, the comparisons in this work are not performed at fixed total energy or fixed external power. Changes in $P_{\mathrm{fus}}$ and $P_{\mathrm{brem}}$ with rotation reflect the combined effects of spatial redistribution, changes in the macroscopic kinetic-energy content, and, when applicable, changes in the adopted drift-Maxwellian reaction-rate coefficient.

\subsubsection*{Fusion-Power Evaluation with a Drift-Maxwellian Reaction-Rate Coefficient}

A relative toroidal flow speed $\Delta u = R |\Omega_p - \Omega_B|$ breaks thermal equilibrium. The fusion reaction-rate coefficient, conventionally termed the fusion reactivity, is calculated with a drift-Maxwellian model~\cite{xie2024upper}. In the present work, this closure is specialized to the near-equal ion-temperature case, $T_p \approx T_B \equiv T_i$, so that the effective thermodynamic temperature reduces to $T_{\mathrm{r}} = T_i$. The corresponding drift kinetic energy is $E_{\mathrm{d}} = \frac{1}{2} m_{\mathrm{r}} (\Delta u)^2$, where $m_{\mathrm{r}}$ is the reduced mass. This treatment is used for power evaluation on top of the calculated equilibria; it does not represent a self-consistent kinetic evolution of the ion distribution functions. Because the closure prescribes a shifted relative-velocity distribution rather than evolving how such a distribution is sustained or collisionally relaxed, it should be read as a collisionless-limit post-processing estimate within the prescribed-profile scan. The resulting non-thermal reaction-rate coefficient $\langle \sigma v \rangle_{\mathrm{DM}}$ is
\begin{equation}
    \langle \sigma v \rangle_{\mathrm{DM}} =
    \sqrt{\frac{2}{\pi m_{\mathrm{r}} k_B^2 T_{\mathrm{r}} E_{\mathrm{d}}}}
    \int_0^\infty \sigma(E)\sqrt{E}
    \exp\left(-\frac{E + E_{\mathrm{d}}}{k_B T_{\mathrm{r}}}\right)
    \sinh\left(\frac{2\sqrt{E E_{\mathrm{d}}}}{k_B T_{\mathrm{r}}}\right)\, dE.
    \label{eq:drift_rate_coefficient}
\end{equation}

For $E_{\mathrm{d}}=0$, Eq.~\eqref{eq:drift_rate_coefficient} is evaluated through its Maxwellian limit, rather than by substituting into the singular written form.

The volume-integrated fusion power is then
\begin{equation}
    P_{\mathrm{fus}} = \int_V n_p(R,Z) n_B(R,Z) \langle \sigma v \rangle_{\mathrm{DM}} \cdot E_{\mathrm{fus}}\, dV,
    \label{eq:fusion_power}
\end{equation}
with $E_{\mathrm{fus}} = 8.7$~MeV.

The EHL-3B zero-dimensional design study in Ref.~\cite{liu2024enn} used updated $p\text{-}^{11}\text{B}$ fusion cross-section data and additionally introduced a potential reaction-rate enhancement factor $f_r$, with $f_r=2$ for its EHL-3B reference point, to represent possible non-Maxwellian or other effects. No such additional global multiplier is applied in the present calculations. Instead, $\langle\sigma v\rangle$ is evaluated directly from $\sigma(E)$ using either the Maxwellian limit or the drift-Maxwellian distribution in Eq.~\eqref{eq:drift_rate_coefficient}. Thus, all reported $P_{\mathrm{fus}}$ and $\mathcal{R}_{\mathrm{fb}}$ values contain the explicitly modeled distribution effect but no additional phenomenological reaction-rate factor.

\subsubsection*{Relativistic Bremsstrahlung Evaluation}

Bremsstrahlung is calculated with the \texttt{fgfit} module~\cite{xie2024bremsstrahlung}. The local power density $p_{\mathrm{br}}$ is integrated over the plasma volume. More detailed models are available for non-thermal electron populations~\cite{peysson2008}, but they are not required under the thermal-electron assumption adopted here.

The local bremsstrahlung power density is written as
\begin{equation}
    p_{\mathrm{br}} = C_{\mathrm{br}} n_e \sqrt{T_e} \left[
    \sum_j n_j Z_j^2 g_{\mathrm{ei}}(t,Z_j) + n_e g_{\mathrm{ee}}(t)
    \right],
    \label{eq:p_brem_local}
\end{equation}
and the corresponding global loss is
\begin{equation}
    P_{\mathrm{brem}} = \int_V p_{\mathrm{br}}\, dV
    = \iint_S p_{\mathrm{br}}(R,Z)\cdot 2\pi R\, dR\, dZ,
    \label{eq:P_brem_global}
\end{equation}
where $t = k_B T_e / m_e c^2$, and $g_{\mathrm{ei}}$ and $g_{\mathrm{ee}}$ are thermally averaged Gaunt factors.

\subsubsection*{Total Stored Energy Calculation}

The total stored energy $W_{\mathrm{tot}}$ is taken as the sum of thermal energy $W_{\mathrm{th}}$ and macroscopic kinetic energy $W_{\mathrm{k}}$:
\begin{equation}
    W_{\mathrm{th}} = \int_V \frac{3}{2} k_B \Big( n_e T_e + n_p T_p + n_B T_B \Big)\, dV,
    \label{eq:thermal_energy}
\end{equation}
\begin{equation}
    W_{\mathrm{k}} = \int_V \frac{1}{2} R^2 \Big( m_p n_p \Omega_p^2 + m_B n_B \Omega_B^2 \Big)\, dV,
    \label{eq:kinetic_energy}
\end{equation}
\begin{equation}
    W_{\mathrm{tot}} = W_{\mathrm{th}} + W_{\mathrm{k}}.
    \label{eq:total_energy}
\end{equation}

\clearpage
\section{Equilibrium Spatial Structure}
\label{sec:spatial}

We first examine the two-dimensional spatial structure of the equilibria in order to identify the mechanisms that govern the density, electrostatic-potential, and local power-density distributions. Unless otherwise stated, the results in this section use the EHL-2 geometry listed in Table~\ref{tab:device_parameters}. EHL-3B shows the same qualitative behavior.

\begin{table}[H]
    \centering
    \caption{Baseline physical parameters for the two geometries considered.}
    \label{tab:device_parameters}
    \resizebox{\textwidth}{!}{%
    \begin{tabular}{lccc}
        \toprule
        \textbf{Parameter} & \textbf{Symbol} & \textbf{EHL-2} & \textbf{EHL-3B} \\
        \midrule
        Major radius & $R_0$ & 1.1~m & 3.2~m \\
        Minor radius & $a$ & 0.57~m & 1.9~m \\
        Elongation & $\kappa$ & 2.2 & 2.5 \\
        Triangularity & $\delta$ & 0.50 & 0.60 \\
        Vacuum toroidal field & $B_0$ & 3.0~T & 4.0~T \\
        Plasma current & $I_p$ & 3.0~MA & 25~MA \\
        \midrule
        Proton/boron density (core) & $n_{p0} / n_{B0}$ & $9.0 \times 10^{19} / 1.0 \times 10^{19}\ \mathrm{m^{-3}}$ & $2.3 \times 10^{20} / 2.5 \times 10^{19}\ \mathrm{m^{-3}}$ \\
        Electron/ion temperature (core) & $T_{e0} / T_{i0}$ & 15 / 30~keV & 35 / $1.4\times10^2$~keV \\
        \bottomrule
    \end{tabular}%
    }
\end{table}

\subsection{Iso-Rotation}

We begin with the iso-rotation case, $\Omega_{0,p} = \Omega_{0,B} \equiv \Omega_0$. The centrifugal force $F_c = m_s \Omega_0^2 R$ scales with species mass, so the response of boron is much stronger than that of protons because $m_B/m_p \approx 11$.

\begin{figure}[!p]
    \centering
    \includegraphics[width=0.78\textwidth]{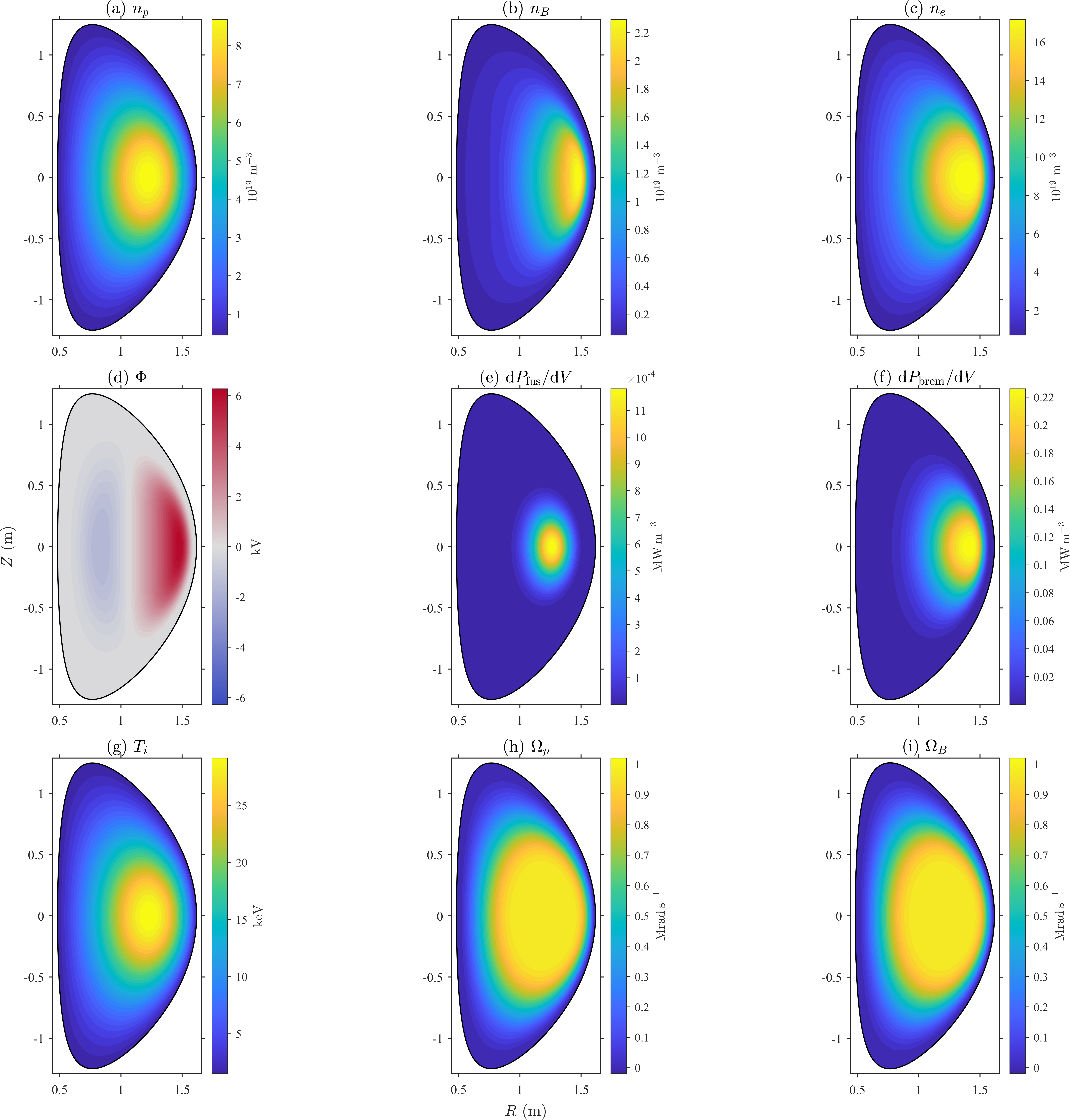}
    \caption{Two-dimensional EHL-2 equilibrium at $\Omega_{p0}=\Omega_{B0}=1.0\times10^6\ \mathrm{rad\,s^{-1}}$. Panels show (a)--(c) proton, boron, and electron densities; (d)--(f) electrostatic potential, local volumetric fusion-power density, and local volumetric bremsstrahlung-power density; and (g)--(i) common ion temperature and proton and boron angular frequencies. The two power-density panels are local integrands; $P_{\mathrm{fus}}$ and $P_{\mathrm{brem}}$ elsewhere denote their volume integrals.}
    \label{fig:2d_profiles}
\end{figure}

Figure~\ref{fig:2d_profiles} shows nine reconstructed fields at $\Omega_0=1.0\times10^6\ \mathrm{rad\,s^{-1}}$. The proton density in panel (a) remains nearly poloidally symmetric, with only a small outward shift. Conversely, the boron density in panel (b) accumulates strongly on the low-field side ($R > R_0$). Since $W_B^{\mathrm{cf}}$ is larger than $W_p^{\mathrm{cf}}$ at the same $R$, the exponential factor in Eq.~\eqref{eq:density_dist} increases $n_B$ more strongly on the outboard side. The high-field side (HFS, $R < R_0$) is correspondingly depleted of boron.

This redistribution would violate local quasi-neutrality without an electrostatic response. Because electrons are light, they do not follow the boron unless an electric field is established. The plasma develops a polarization potential $\Phi$, shown in panel (d), which is positive on the LFS and negative on the HFS. This potential shifts the electron density in panel (c) outward so that $n_e \approx n_p + 5n_B$ is maintained. The local fusion-power density in panel (e) is concentrated where the proton and boron distributions overlap, whereas the bremsstrahlung-power density in panel (f) follows the electron-density redistribution. Panels (g)--(i) show the prescribed common ion-temperature profile and the two angular-frequency profiles used to construct this equilibrium.

The same polarization increases the local electron density on the LFS. Because $n_e$ is shifted outward to compensate the boron redistribution, the volume-integrated $n_e^2$ also increases. The volume element $dV = 2\pi R\, dR\, dZ$ places additional weight on the outboard side and further increases this contribution to global integrals.

\begin{figure}[!p]
    \centering
    \includegraphics[width=0.66\textwidth]{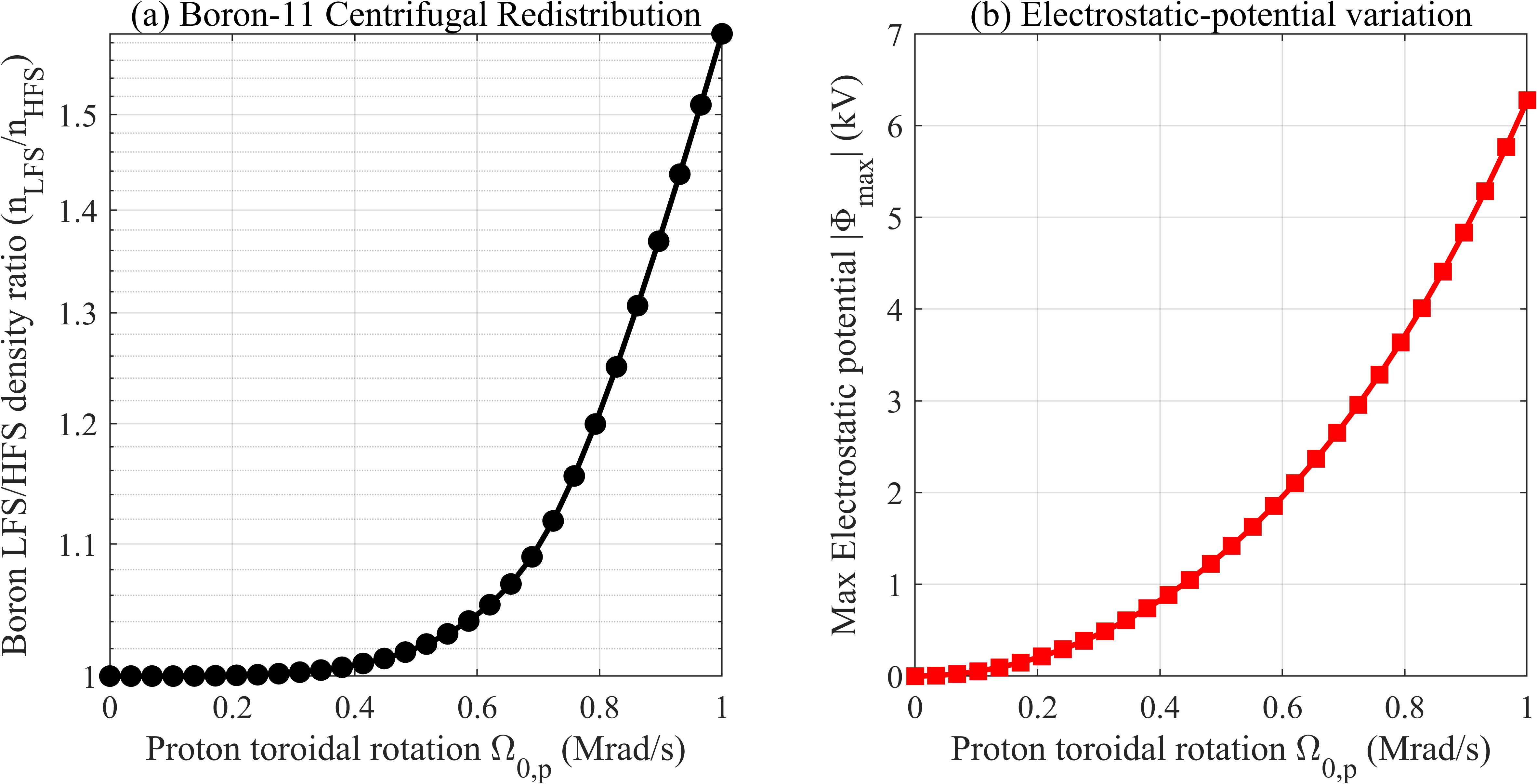}
    \caption{Evolution with $\Omega_0$ under iso-rotation: (a) boron density asymmetry and (b) electrostatic-potential well depth.}
    \label{fig:1d_mechanisms}
\end{figure}

Figure~\ref{fig:1d_mechanisms} shows how this polarization scales with $\Omega_0$. The boron density asymmetry $n_B^{\text{LFS}} / n_B^{\text{HFS}}$ in panel (a) grows approximately exponentially with $\Omega_0^2$. This follows from the ratio of the Boltzmann factors at two radial locations, where the centrifugal-potential difference is proportional to $\Omega_0^2$. The potential-well depth in panel (b) increases accordingly.

\FloatBarrier

\subsection{Differential Toroidal Rotation}

\begin{figure}[H]
    \centering
    \includegraphics[width=0.76\textwidth]{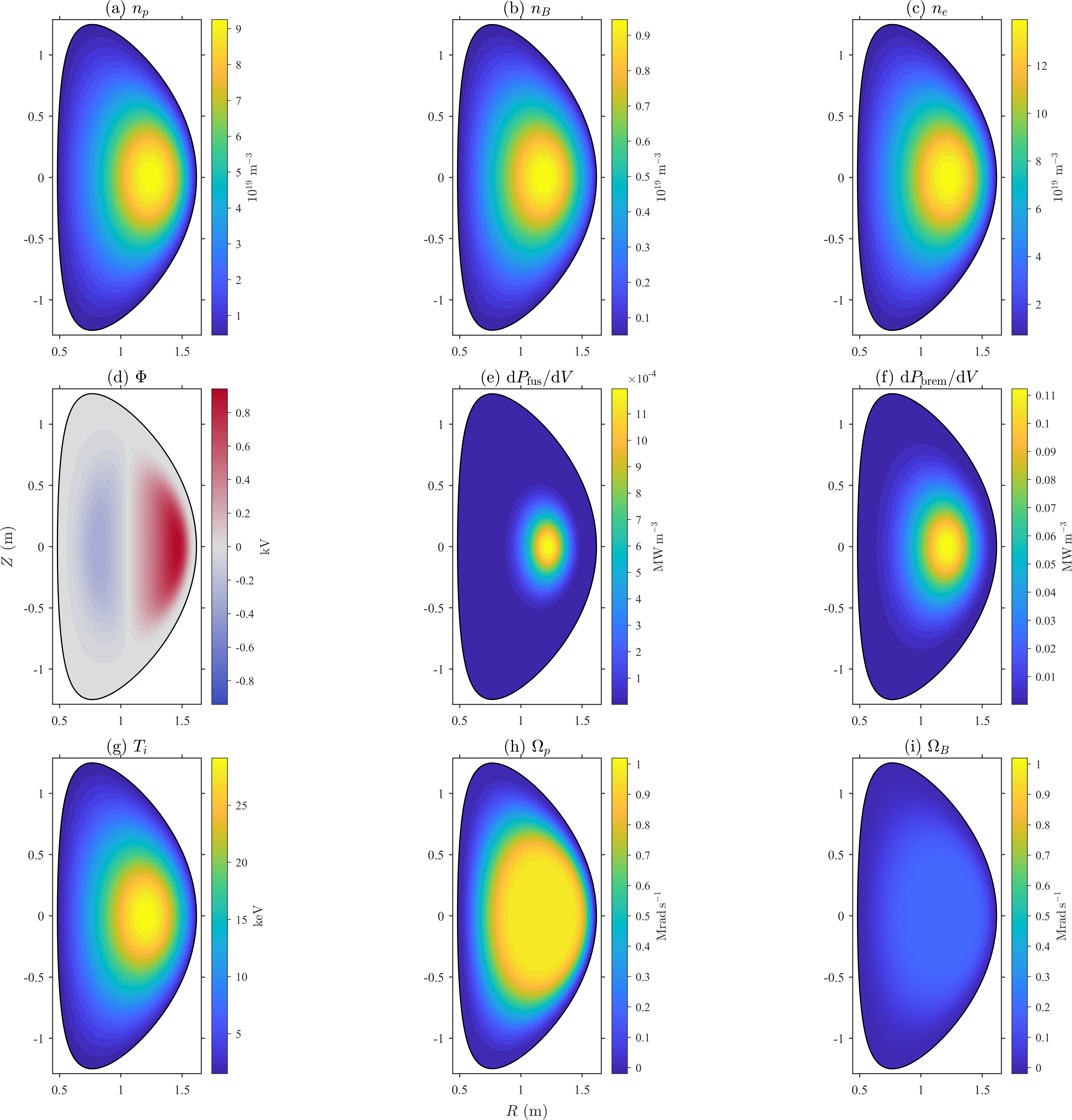}
    \caption{Two-dimensional EHL-2 equilibrium at $\Omega_{p0}=1.0\times10^6\ \mathrm{rad\,s^{-1}}$ and $\gamma_{Bp}=\Omega_{B0}/\Omega_{p0}=0.20$. Panels show (a)--(c) proton, boron, and electron densities; (d)--(f) electrostatic potential, local volumetric fusion-power density, and local volumetric bremsstrahlung-power density; and (g)--(i) common ion temperature and proton and boron angular frequencies. Panels (h) and (i) use a common color scale.}
    \label{fig:diff_2d_profiles}
\end{figure}

Figure~\ref{fig:diff_2d_profiles} shows the same nine fields for a fixed rotation ratio $\gamma_{Bp}=\Omega_{B0}/\Omega_{p0}=0.20$. Relative to the iso-rotation case, the boron accumulation on the LFS is reduced and the polarization potential is shallower.

Reducing $\Omega_B$ lowers $W_B^{\mathrm{cf}} \propto \Omega_B^2$ and weakens the exponential drive that pushes boron toward the LFS. The boron density becomes more symmetric. Because less boron accumulates on the LFS, a smaller polarization potential $\Phi$ is required, and the outward shift of electrons is also reduced. The electron density profile becomes flatter. Consistent with the $n_e^2$ dependence of bremsstrahlung, panel (f) shows a lower and less outboard-localized bremsstrahlung-power density than in the iso-rotation case.

\section{\texorpdfstring{$\mathcal{R}_{\mathrm{fb}}$}{Rfb} Parameter Scans}
\label{sec:parameter}

This section examines how $P_{\mathrm{fus}}$, $P_{\mathrm{brem}}$, and $\mathcal{R}_{\mathrm{fb}}=P_{\mathrm{fus}}/P_{\mathrm{brem}}$ vary across the scanned parameter space for EHL-2 and EHL-3B. The core rotation frequencies $\Omega_{p0}$ and $\Omega_{B0}$ are varied, while the on-axis amplitudes and shape parameters of the density and temperature profiles are kept at the baseline values in Table~\ref{tab:device_parameters} and Appendix~\ref{app:profile_sensitivity}. The scans are therefore performed at fixed prescribed profile amplitudes rather than at fixed total particle inventories. Variations in the global power integrals can accordingly reflect both spatial redistribution and changes in the total species inventories implied by the equilibrium solution.

\subsection{Iso-Rotation Scans}

\begin{figure}[!htbp]
    \centering
    \includegraphics[width=0.8\textwidth]{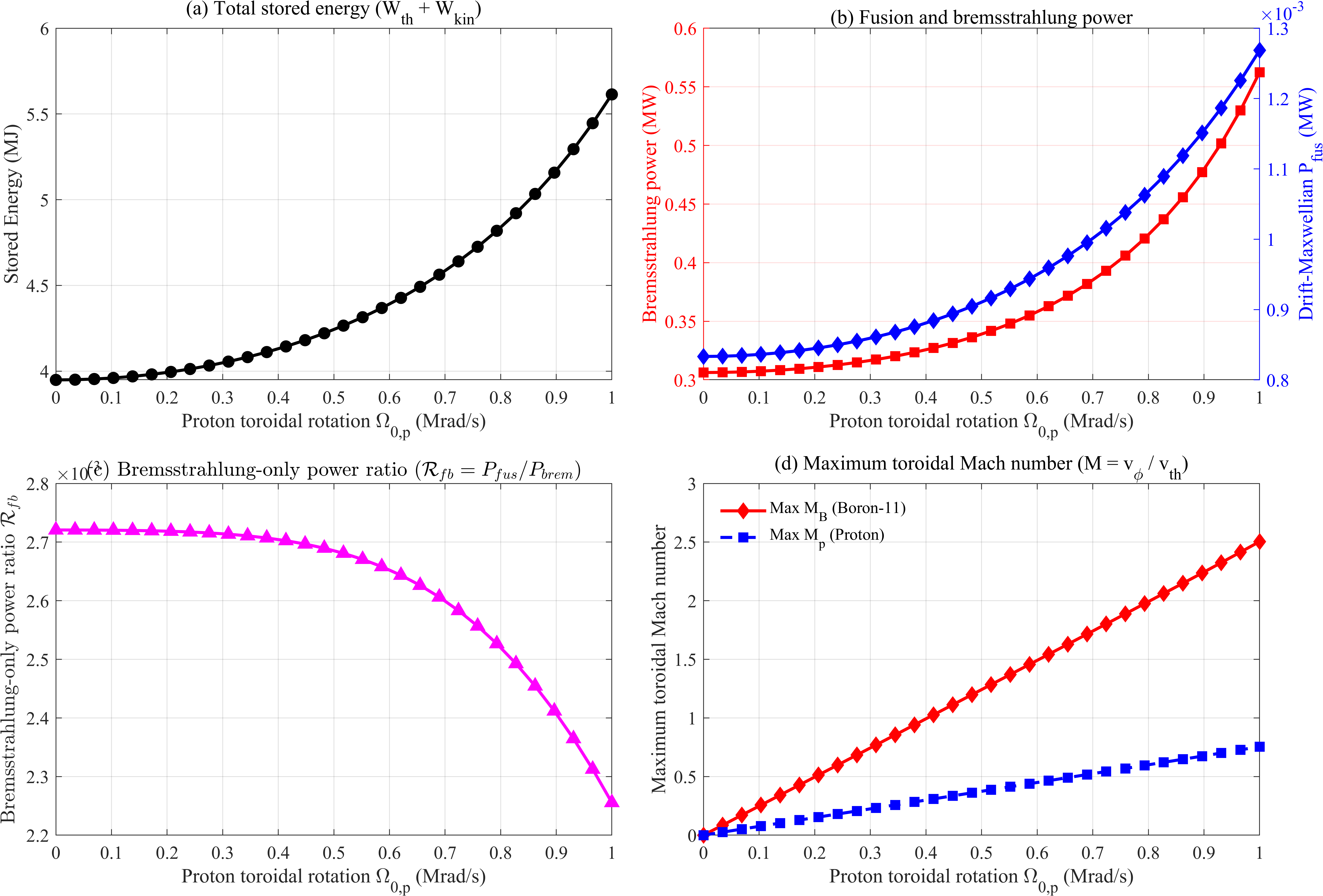}
    \caption{Iso-rotation scan for EHL-2: (a) total stored energy, (b) $P_{\mathrm{fus}}$ and $P_{\mathrm{brem}}$, (c) $\mathcal{R}_{\mathrm{fb}}$, and (d) Mach numbers.}
    \label{fig:scan_metrics}
\end{figure}

Figure~\ref{fig:scan_metrics} shows the macroscopic quantities along the iso-rotation line $\Omega_{0,p} = \Omega_{0,B} = \Omega_0$ for EHL-2. The total stored energy $W_{\mathrm{tot}}$ in panel (a) increases with $\Omega_0$, mainly because $W_{\mathrm{k}} \propto \int n_i m_i \Omega_0^2 R^2\, dV$. The thermal part $W_{\mathrm{th}}$ changes much less because the temperature profiles are fixed.

Panel (b) compares $P_{\mathrm{fus}}$ and $P_{\mathrm{brem}}$. Both increase with $\Omega_0$, but $P_{\mathrm{brem}}$ grows more rapidly. Under iso-rotation, $P_{\mathrm{fus}}=E_{\mathrm{fus}}\int n_p n_B\langle\sigma v\rangle_{\mathrm{M}}(T_i)\,dV$ contains the bilinear proton--boron density overlap weighted by the prescribed Maxwellian reaction-rate-coefficient profile, whereas $P_{\mathrm{brem}}$ is governed primarily by the quadratic electron-density integral. Centrifugal accumulation on the LFS increases $n_B$ and $n_e$ locally, and the $n_e^2$ dependence amplifies the change in bremsstrahlung. The LFS volume weighting strengthens this effect further.

Consequently, $\mathcal{R}_{\mathrm{fb}}$ in panel (c) decreases with increasing $\Omega_0$. At $\Omega_0 = 0$, $\mathcal{R}_{\mathrm{fb}} \approx 0.40$, while at $\Omega_0 = 1.0\times 10^6$~rad/s it falls below $0.10$. Panel (d) shows the Mach numbers. Since $M_s \propto \Omega_0 / \sqrt{T_s/m_s}$, the heavier boron species has the larger Mach number, and both species become supersonic at sufficiently large $\Omega_0$.

The corresponding iso-rotation scan for EHL-3B is shown in Fig.~\ref{fig:hl3b_compare}(a). The larger major radius, $R_0 = 3.2$~m, increases the centrifugal effect, so the accessible scan range in $\Omega_0$ is chosen to be lower than that used for EHL-2 in order to remain within the regime of well-behaved equilibria (see Appendix~\ref{app:rotation_limit}). At $\Omega_0=0.90$~Mrad/s, $P_{\mathrm{brem}}$ reaches about $1.6\times10^3$~MW while $P_{\mathrm{fus}}$ is about $2.9\times10^2$~MW, so $\mathcal{R}_{\mathrm{fb}}$ falls to $0.18$ along the iso-rotation line. Within the present EHL-3B model case, strong iso-rotation therefore increases bremsstrahlung.

\subsection{Differential Toroidal Rotation Scans}

\begin{figure}[!htbp]
    \centering
    \includegraphics[width=0.8\textwidth]{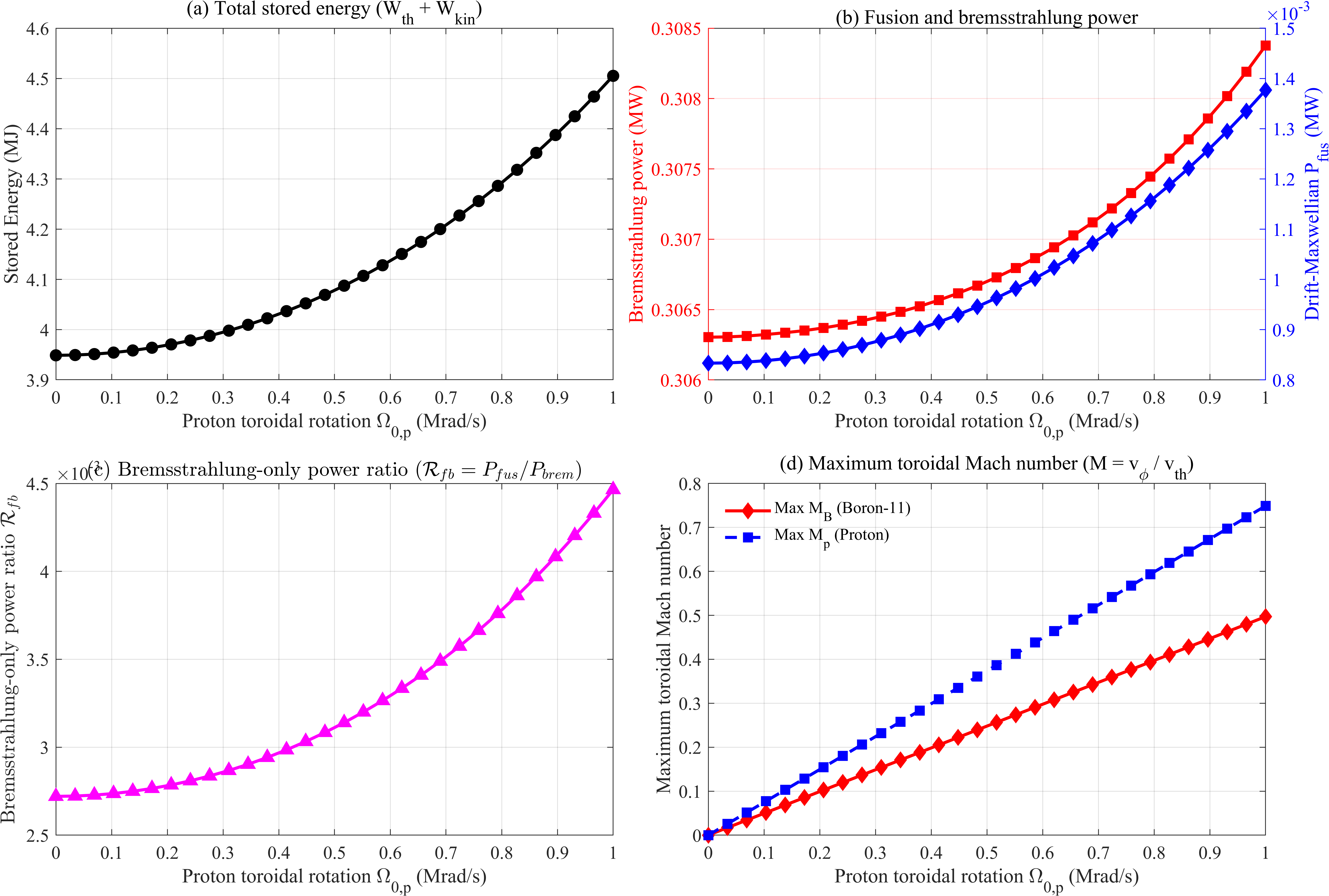}
    \caption{Species-dependent toroidal-rotation scan ($\gamma_{Bp}=0.20$) for EHL-2: (a) total stored energy, (b) $P_{\mathrm{fus}}$ and $P_{\mathrm{brem}}$, and (c) $\mathcal{R}_{\mathrm{fb}}$.}
    \label{fig:diff_scan_metrics}
\end{figure}

Figure~\ref{fig:diff_scan_metrics} shows the EHL-2 results for a fixed ratio $\gamma_{Bp}=0.20$. As $\Omega_{p0}$ increases, $P_{\mathrm{brem}}$ grows sub-linearly, unlike the iso-rotation case. This behavior follows from the reduced centrifugal polarization discussed in Section~\ref{sec:spatial}. With $\Omega_{B0}=0.20\Omega_{p0}$, the boron accumulation on the LFS is weaker, the electron density profile is flatter, and the integral of $n_e^2$ is correspondingly reduced.

The fusion power $P_{\mathrm{fus}}$ also increases more slowly than in the iso-rotation case. As a consequence, $\mathcal{R}_{\mathrm{fb}}$ rises only modestly with $\Omega_{p0}$. For the smaller major radius of EHL-2, the relative toroidal flow produces only a limited drift kinetic energy ($E_{\mathrm{d}} < 5.0$~keV), so the change in the drift-Maxwellian reaction-rate coefficient is negligible. The increase in $\mathcal{R}_{\mathrm{fb}}$ is associated mainly with the reduction in bremsstrahlung.

\begin{figure}[!htbp]
    \centering
    \begin{subfigure}{0.48\textwidth}
        \centering
        \includegraphics[width=\linewidth]{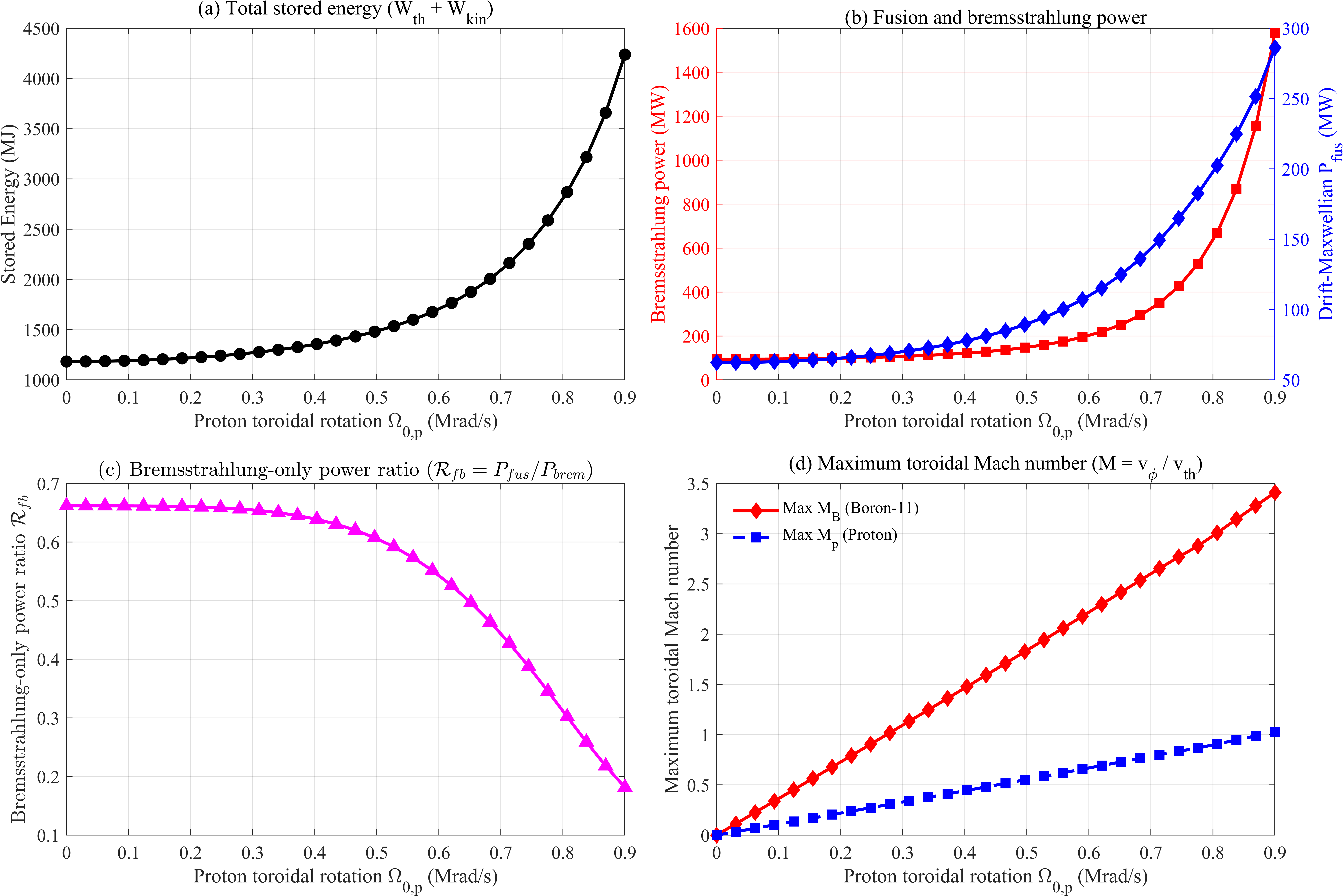}
        \caption{Iso-rotation ($\gamma_{Bp}=1.0$)}
    \end{subfigure}\hfill
    \begin{subfigure}{0.48\textwidth}
        \centering
        \includegraphics[width=\linewidth]{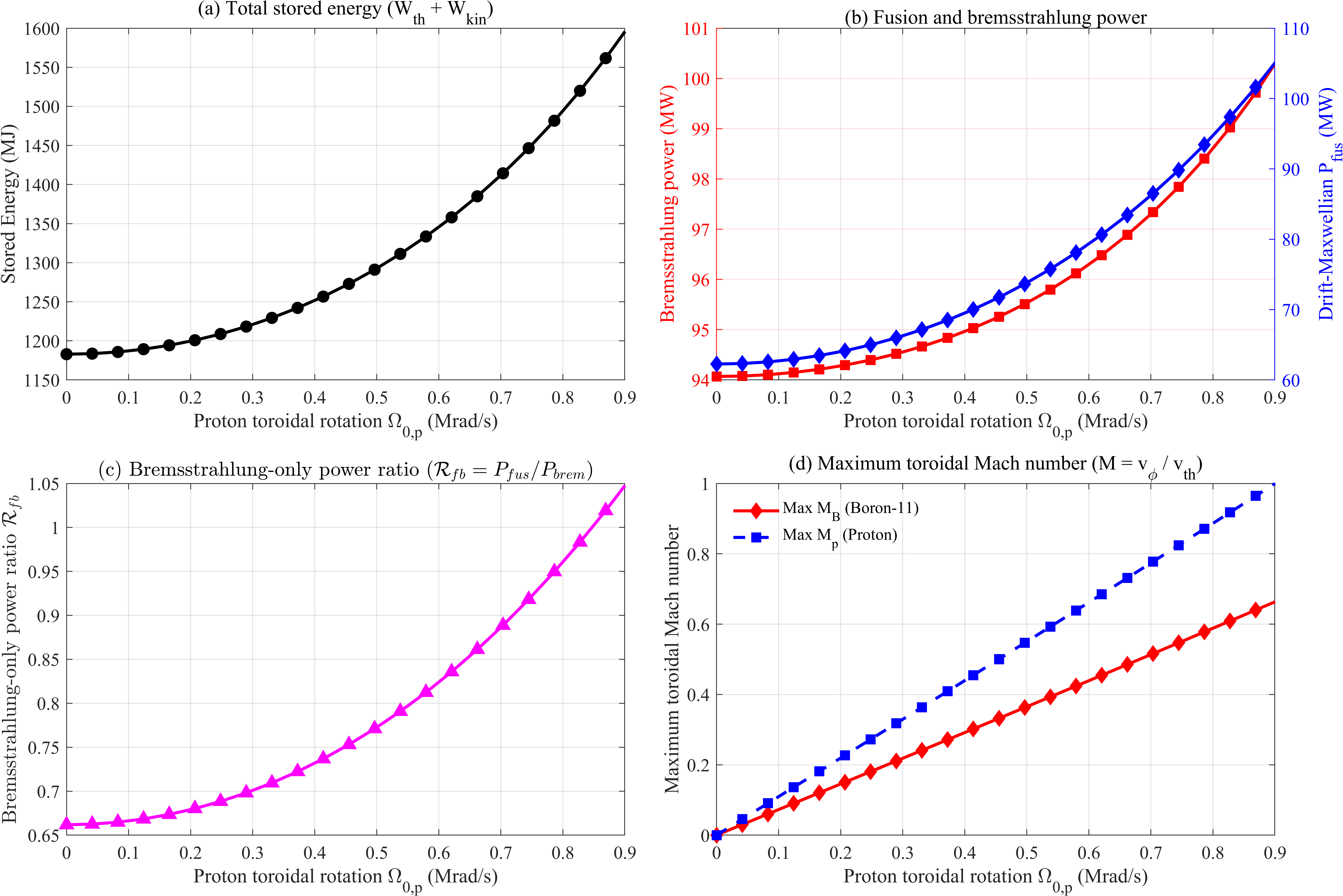}
        \caption{Species-dependent rotation ($\gamma_{Bp}=0.20$)}
    \end{subfigure}
    \caption{EHL-3B rotation-strength scans. Panel (a) shows the iso-rotation line, where increasing $\Omega_0$ raises $P_{\mathrm{brem}}$ and lowers $\mathcal{R}_{\mathrm{fb}}$. Panel (b) shows the species-dependent toroidal-rotation scan with $\gamma_{Bp}=0.20$, where $P_{\mathrm{brem}}$ remains near $1.0\times10^2$~MW and the bremsstrahlung-only power ratio $\mathcal{R}_{\mathrm{fb}}$ reaches the $\mathcal{R}_{\mathrm{fb}}=1$ crossing level near the upper end of the reported scan range.}
    \label{fig:hl3b_compare}
\end{figure}

For EHL-3B, the relative toroidal flow has a stronger effect. Figure~\ref{fig:hl3b_compare}(b) shows the species-dependent toroidal-rotation scan for $\gamma_{Bp}=0.20$ over the reported range $0 \leq \Omega_{p0} \leq 0.90$~Mrad/s. As in EHL-2, $P_{\mathrm{brem}}$ remains close to $1.0\times10^2$~MW across the scanned $\Omega_{p0}$ range. In contrast, $P_{\mathrm{fus}}$ rises steadily with $\Omega_{p0}$ and approaches the same order as $P_{\mathrm{brem}}$ near the upper end of the scan.

For comparison, evaluating the zero-dimensional EHL-3B model with $f_r=1$ gives $P_{\mathrm{fus}}\approx6.7\times10^1$~MW and $P_{\mathrm{brem}}\approx9.1\times10^1$~MW, close to the non-rotating VEQ-MF values of approximately $6.2\times10^1$~MW and $9.4\times10^1$~MW, respectively. The larger $P_{\mathrm{fus}}\approx1.3\times10^2$~MW quoted for the roadmap EHL-3B point includes the $f_r=2$ scenario multiplier and is therefore not the direct baseline for the present calculations.

This increase in fusion power traces to the drift kinetic energy
\[
E_{\mathrm{d}} = \frac{1}{2} m_{\mathrm{r}} R^2 (\Omega_p - \Omega_B)^2.
\]
For EHL-3B, the combination of large major radius and large $\Delta \Omega$ gives tens of keV of drift kinetic energy, which is sufficient to raise $\langle\sigma v\rangle_{\mathrm{DM}}$ through the drift-Maxwellian mechanism. Together with the density redistribution, this increase strengthens fusion power relative to bremsstrahlung and brings the bremsstrahlung-only power ratio $\mathcal{R}_{\mathrm{fb}}$ to the $\mathcal{R}_{\mathrm{fb}}=1$ crossing level near the upper end of the fixed-$\gamma_{Bp}$ scan.

\begin{figure}[!htbp]
    \centering
    \includegraphics[width=0.9\textwidth]{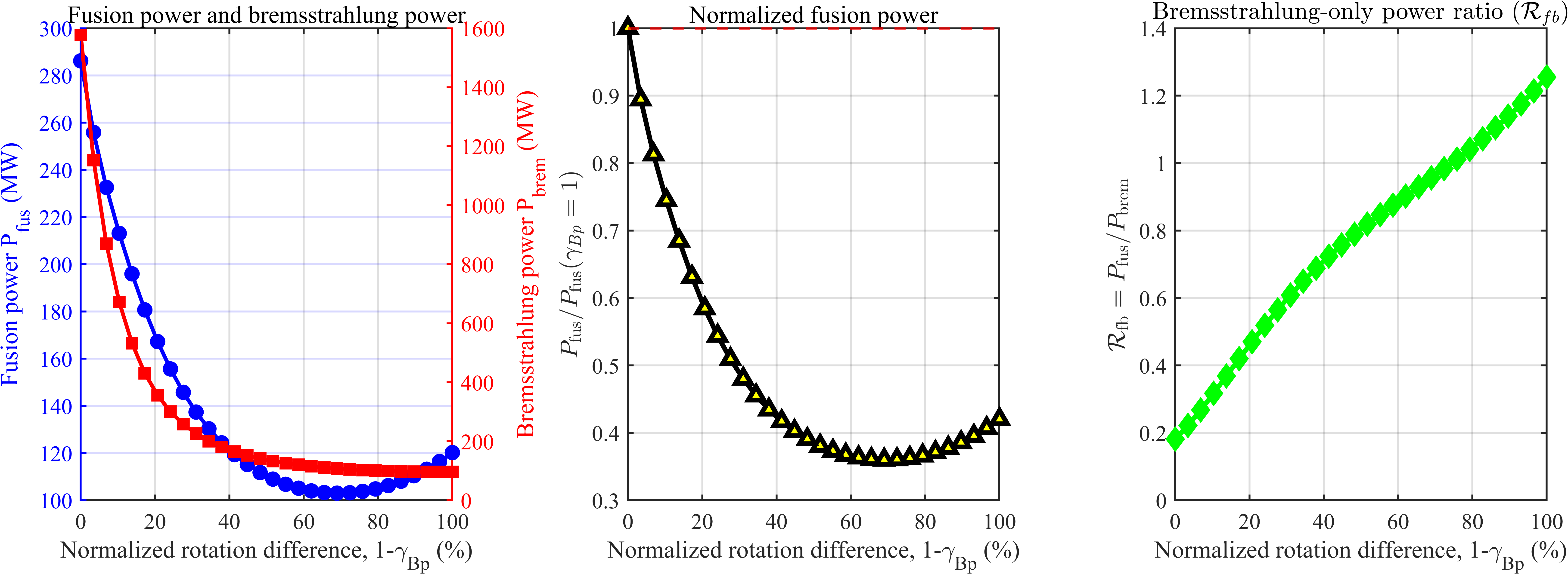}
    \caption{EHL-3B scan over the rotation ratio $\gamma_{Bp}=\Omega_{B0}/\Omega_{p0}$ at fixed $\Omega_{p0}=0.90$~Mrad/s. As $\gamma_{Bp}$ decreases from unity to zero, $P_{\mathrm{brem}}$ falls from about $1.6\times10^3$ to $9.6\times10^1$~MW, while $P_{\mathrm{fus}}$ first decreases and then recovers to about $1.2\times10^2$~MW. The resulting bremsstrahlung-only power ratio $\mathcal{R}_{\mathrm{fb}}$ rises from 0.18 to about 1.3.}
    \label{fig:hl3b_gamma_scan}
\end{figure}

Figure~\ref{fig:hl3b_gamma_scan} shows the continuous dependence on $\gamma_{Bp}$ at fixed $\Omega_{p0}$. As $\gamma_{Bp}$ decreases from unity, both $P_{\mathrm{fus}}$ and $P_{\mathrm{brem}}$ initially decrease. This is the signature of reduced centrifugal polarization: reducing $\Omega_B$ flattens the density profiles and lowers both density integrals. The diagnostic decomposition in Fig.~\ref{fig:mechanism_inventory} shows that the bremsstrahlung density integral $\int n_e^2\,dV$ decreases more strongly than the proton--boron overlap integral $\int n_pn_B\,dV$ over this part of the scan. The full fusion power remains $E_{\mathrm{fus}}\int n_pn_B\langle\sigma v\rangle_{\mathrm{DM}}\,dV$; Section~\ref{sec:mechanisms} separates the density-overlap and reaction-rate-coefficient contributions to this power. Consequently, $P_{\mathrm{brem}}$ initially decreases faster and $\mathcal{R}_{\mathrm{fb}}$ begins to rise.

For $\gamma_{Bp} \lesssim 0.30$, however, $P_{\mathrm{fus}}$ recovers from its intermediate-$\gamma_{Bp}$ minimum. This recovery is driven by the increase in $\langle\sigma v\rangle_{\mathrm{DM}}$, which offsets part of the declining proton--boron overlap. As $\gamma_{Bp}$ decreases, $\Delta\Omega$ increases, $E_{\mathrm{d}}$ grows, and the $\sinh$ term in Eq.~\eqref{eq:drift_rate_coefficient} increases the high-energy tail of the relative-velocity distribution. In the present fixed-$\Omega_{p0}=0.90$~Mrad/s scan, $P_{\mathrm{fus}}$ decreases from about $2.9\times10^2$~MW at $\gamma_{Bp}=1$ to a minimum of about $1.0\times10^2$~MW near $\gamma_{Bp}=0.31$, and then recovers to about $1.2\times10^2$~MW at $\gamma_{Bp}=0$. At the same time, $P_{\mathrm{brem}}$ continues to fall from about $1.6\times10^3$~MW to $9.6\times10^1$~MW. The net result is a steady rise in the bremsstrahlung-only power ratio $\mathcal{R}_{\mathrm{fb}}$ from $0.18$ to about $1.3$.

The latter value should be read together with the omitted loss and sustainment scales. For the same largest imposed relative-toroidal-flow EHL-3B point, the post-processed powers are approximately $P_{\mathrm{fus}}\simeq1.2\times10^2$~MW and $P_{\mathrm{brem}}\simeq9.6\times10^1$~MW, while Appendix~\ref{app:momentum_exchange} gives an order-of-magnitude momentum-exchange power scale of $P_{\mathrm{fric}}\simeq5.2$~GW. This estimate is not added to the denominator of $\mathcal{R}_{\mathrm{fb}}$; it instead emphasizes that the crossing discussed here belongs to the reduced bremsstrahlung-only post-processing indicator.

The fixed-$\gamma_{Bp}$ scan in Fig.~\ref{fig:hl3b_compare}(b) and the fixed-$\Omega_{p0}$ scan in Fig.~\ref{fig:hl3b_gamma_scan} overlap near $\Omega_{p0}\simeq0.90$~Mrad/s and $\gamma_{Bp}\simeq0.20$. Re-evaluating the existing scan caches gives values near $\mathcal{R}_{\mathrm{fb}}\simeq1.0$ on both paths, with a relative difference of about $1.5\%$. This agreement provides a cross-check between the two scan paths at the overlapping point; it is not an independent solver benchmark.

\subsection{Two-Dimensional Maps of \texorpdfstring{$\mathcal{R}_{\mathrm{fb}}$}{Rfb}}

\begin{figure}[!htbp]
    \centering
    \begin{subfigure}{0.48\textwidth}
        \centering
        \includegraphics[width=\linewidth]{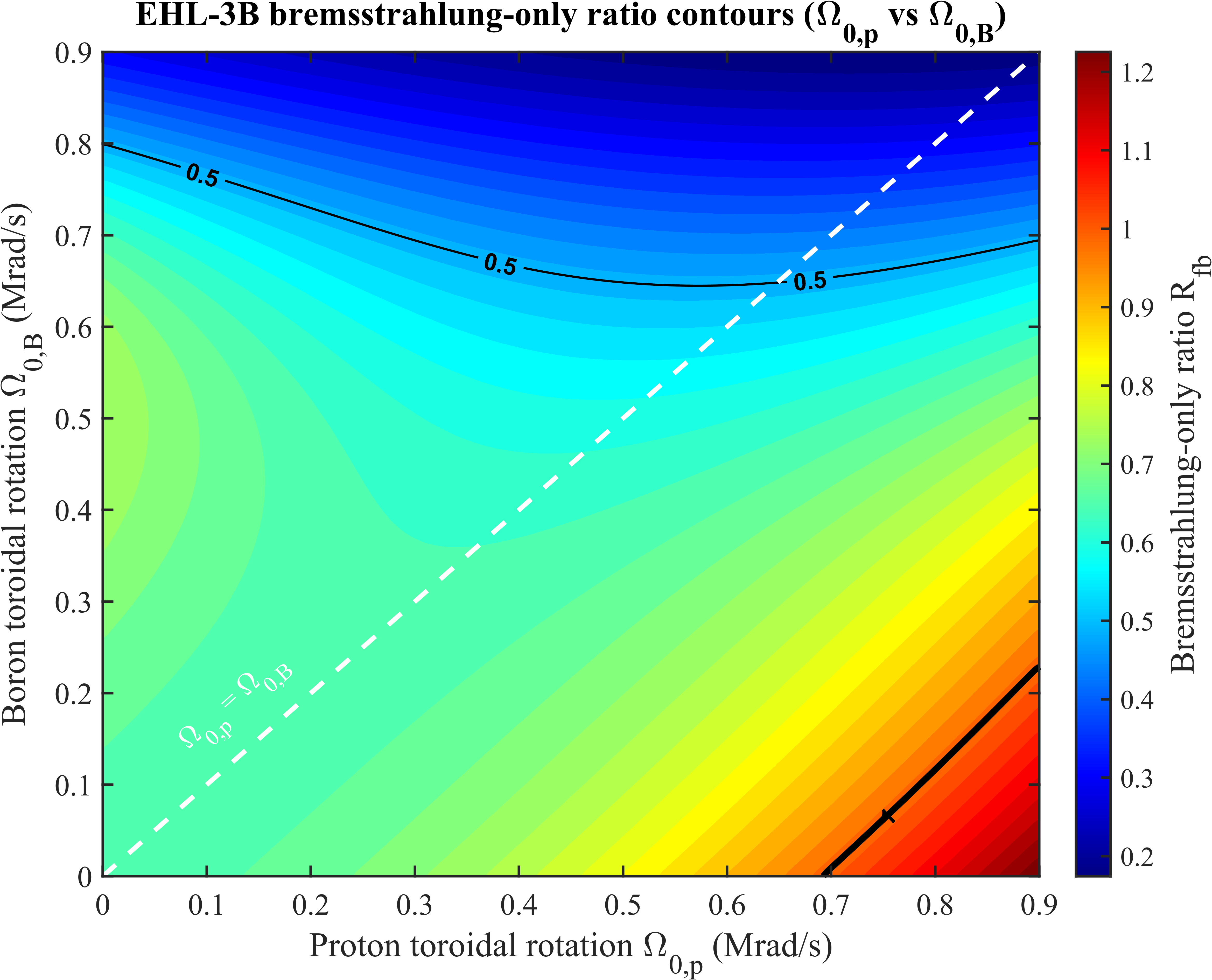}
        \caption{$\Omega_{0,B}$ versus $\Omega_{0,p}$}
    \end{subfigure}\hfill
    \begin{subfigure}{0.48\textwidth}
        \centering
        \includegraphics[width=\linewidth]{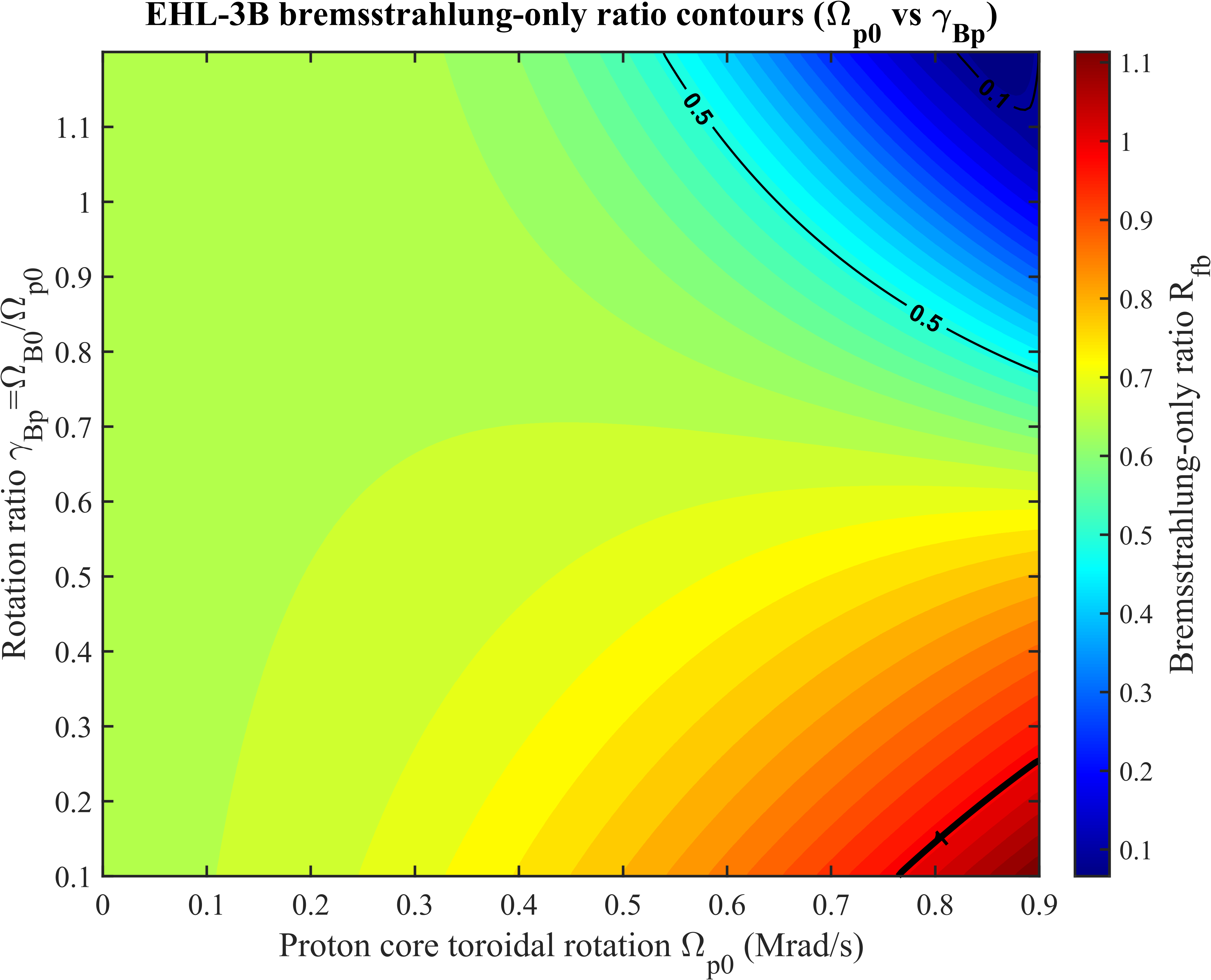}
        \caption{$\gamma_{Bp}$ versus $\Omega_{p0}$}
    \end{subfigure}
    \caption{$\mathcal{R}_{\mathrm{fb}}$ contours for EHL-3B. Panel (a) is shown in absolute-frequency space, and panel (b) in rotation-ratio space. The black contour marks $\mathcal{R}_{\mathrm{fb}}=1$ for the bremsstrahlung-only power ratio; it is not a reactor-gain contour or a full loss-balance contour.}
    \label{fig:popcon_rotation}
\end{figure}

To identify where the largest bremsstrahlung-only $\mathcal{R}_{\mathrm{fb}}$ values occur in the scanned space, two-dimensional scans over the $(\Omega_{0,p}, \Omega_{0,B})$ plane are constructed. Figure~\ref{fig:popcon_rotation} shows the corresponding contours for EHL-3B. The white dashed line denotes iso-rotation. Along this line, $\mathcal{R}_{\mathrm{fb}}$ remains well below unity over the scan.

The largest $\mathcal{R}_{\mathrm{fb}}$ values appear in the lower-right part of Fig.~\ref{fig:popcon_rotation}(a), where $\Omega_{p0}$ is large and $\Omega_{B0}$ is small, corresponding to large relative toroidal flow. In Fig.~\ref{fig:popcon_rotation}(b), the solid black contour marks $\mathcal{R}_{\mathrm{fb}}=1$ for this bremsstrahlung-only power ratio. Within the present scan, this contour appears only in the high-$\Omega_{p0}$, low-$\gamma_{Bp}$ part of the model parameter space.

Although the numerical error near the high-rotation boundary is expected to be larger than that in the lower-parameter region, the in-range benchmark results in Table~\ref{tab:in_range_benchmark} show few-percent FDM/VEQ-MF agreement in the global quantities used for the power-ratio comparison. The representative modal and grid checks in Table~\ref{tab:convergence_in_range} further show small VEQ-MF sensitivity at $\Omega_{p0}=0.90$~Mrad/s and $\gamma_{Bp}=0.20$. The $\mathcal{R}_{\mathrm{fb}}=1$ contour in Fig.~\ref{fig:popcon_rotation} is therefore a numerically resolved bremsstrahlung-only contour within the present model and discretization. Its robustness as a sharp model boundary still requires broader testing.

Here $\mathcal{R}_{\mathrm{fb}}=1$ means that fusion power equals bremsstrahlung loss within the present post-processing power ratio; other radiative and transport losses, auxiliary power, and sustainment requirements are not included. Reaching this bremsstrahlung-only crossing level does not imply reactor-level net-energy feasibility. It shows that, relative to the iso-rotation case, the imposed relative toroidal flow shifts the fusion-to-bremsstrahlung balance toward fusion in this reduced calculation.

\begin{figure}[!htbp]
    \centering
    \begin{subfigure}{0.48\textwidth}
        \centering
        \includegraphics[width=\linewidth]{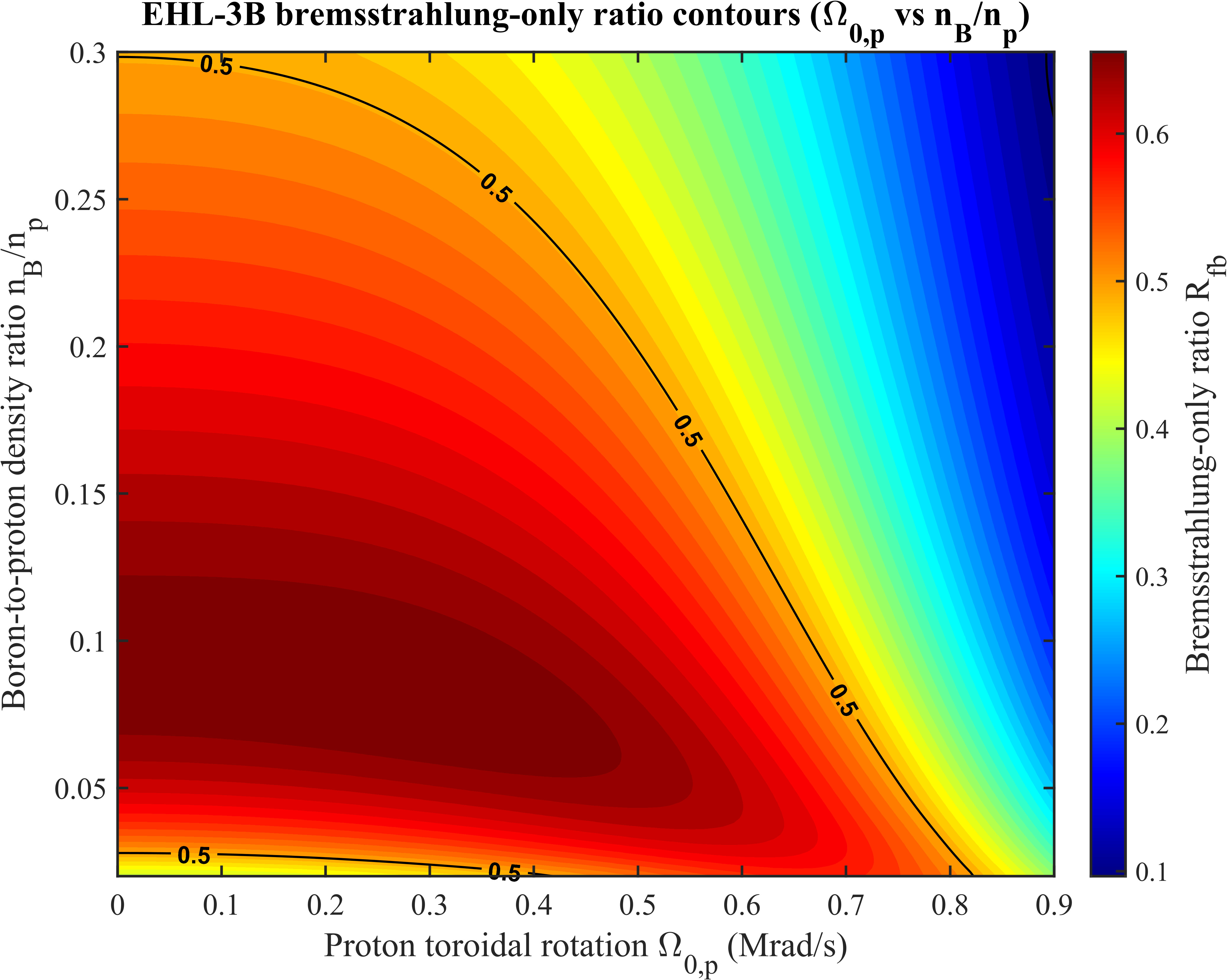}
        \caption{Maxwellian fusion-power evaluation}
    \end{subfigure}\hfill
    \begin{subfigure}{0.48\textwidth}
        \centering
        \includegraphics[width=\linewidth]{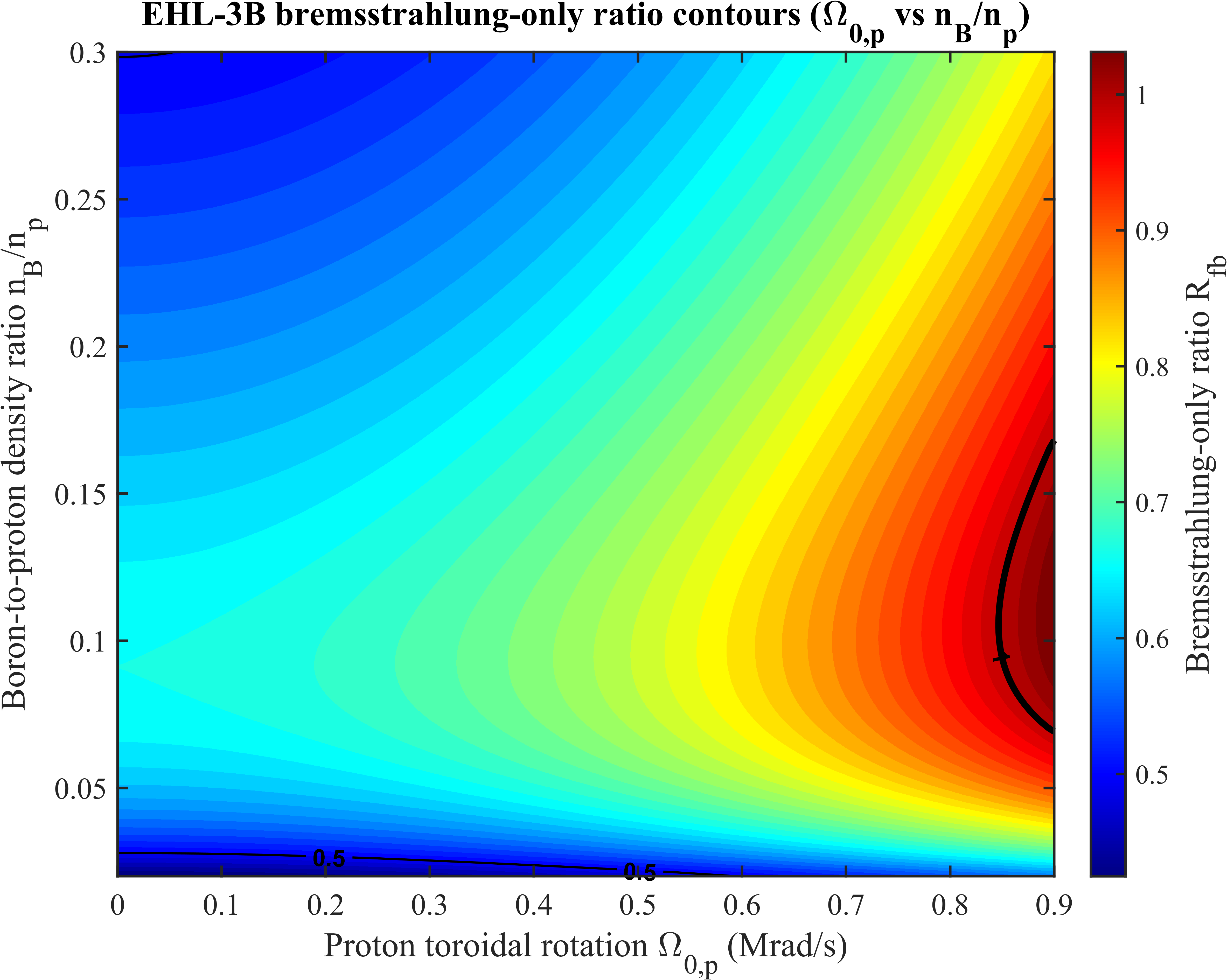}
        \caption{Drift-Maxwellian fusion-power evaluation}
    \end{subfigure}
    \caption{$\mathcal{R}_{\mathrm{fb}}$ as a function of $n_B/n_p$ and $\Omega_{0,p}$ for EHL-3B. The thermal reference remains below unity, whereas the drift-Maxwellian post-processing gives bremsstrahlung-only $\mathcal{R}_{\mathrm{fb}}>1$ values at sufficiently large proton rotation and intermediate boron fraction.}
    \label{fig:popcon_fuel}
\end{figure}

\begin{figure}[!htbp]
    \centering
    \begin{subfigure}{0.48\textwidth}
        \centering
        \includegraphics[width=\linewidth]{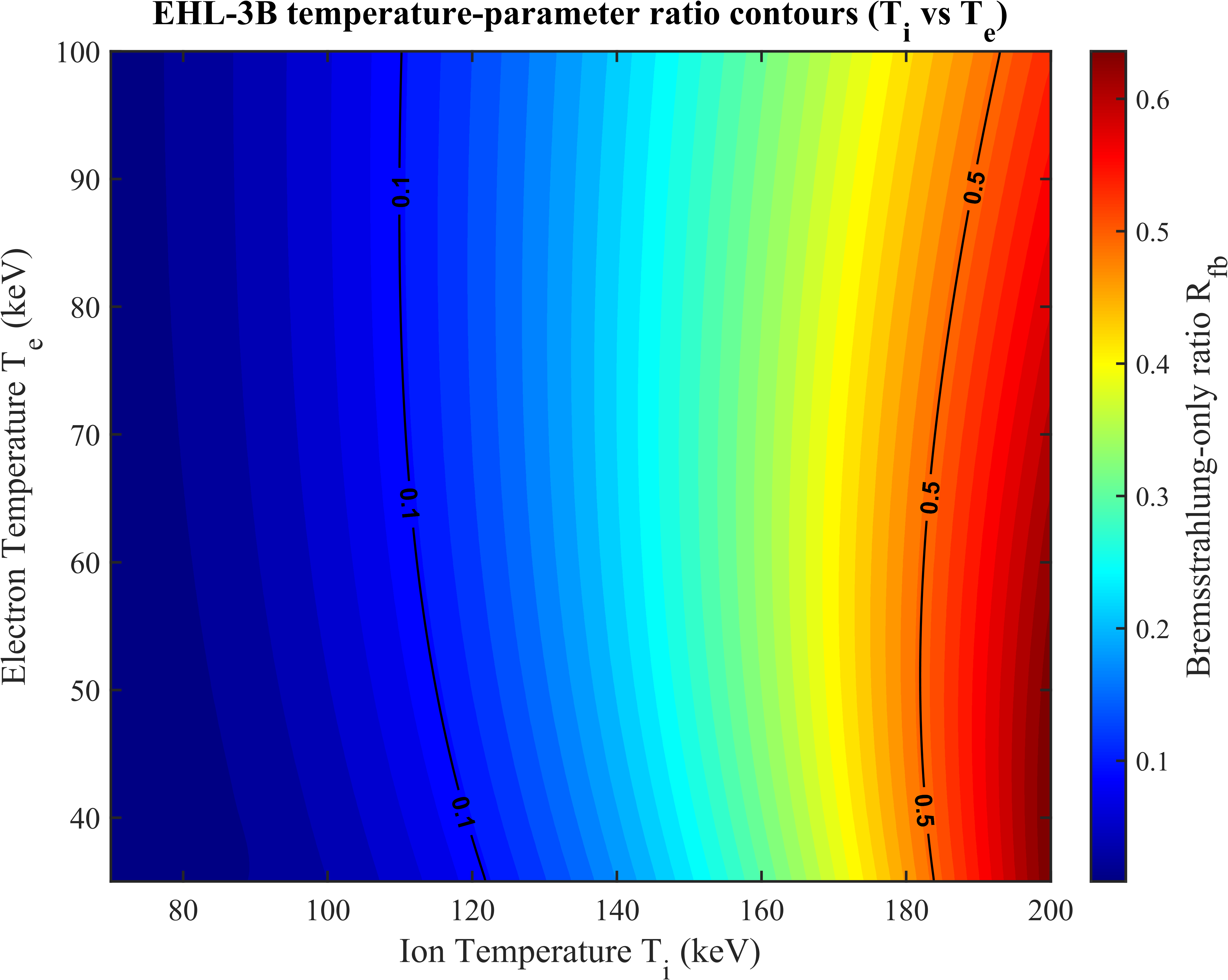}
        \caption{Maxwellian fusion-power evaluation}
    \end{subfigure}\hfill
    \begin{subfigure}{0.48\textwidth}
        \centering
        \includegraphics[width=\linewidth]{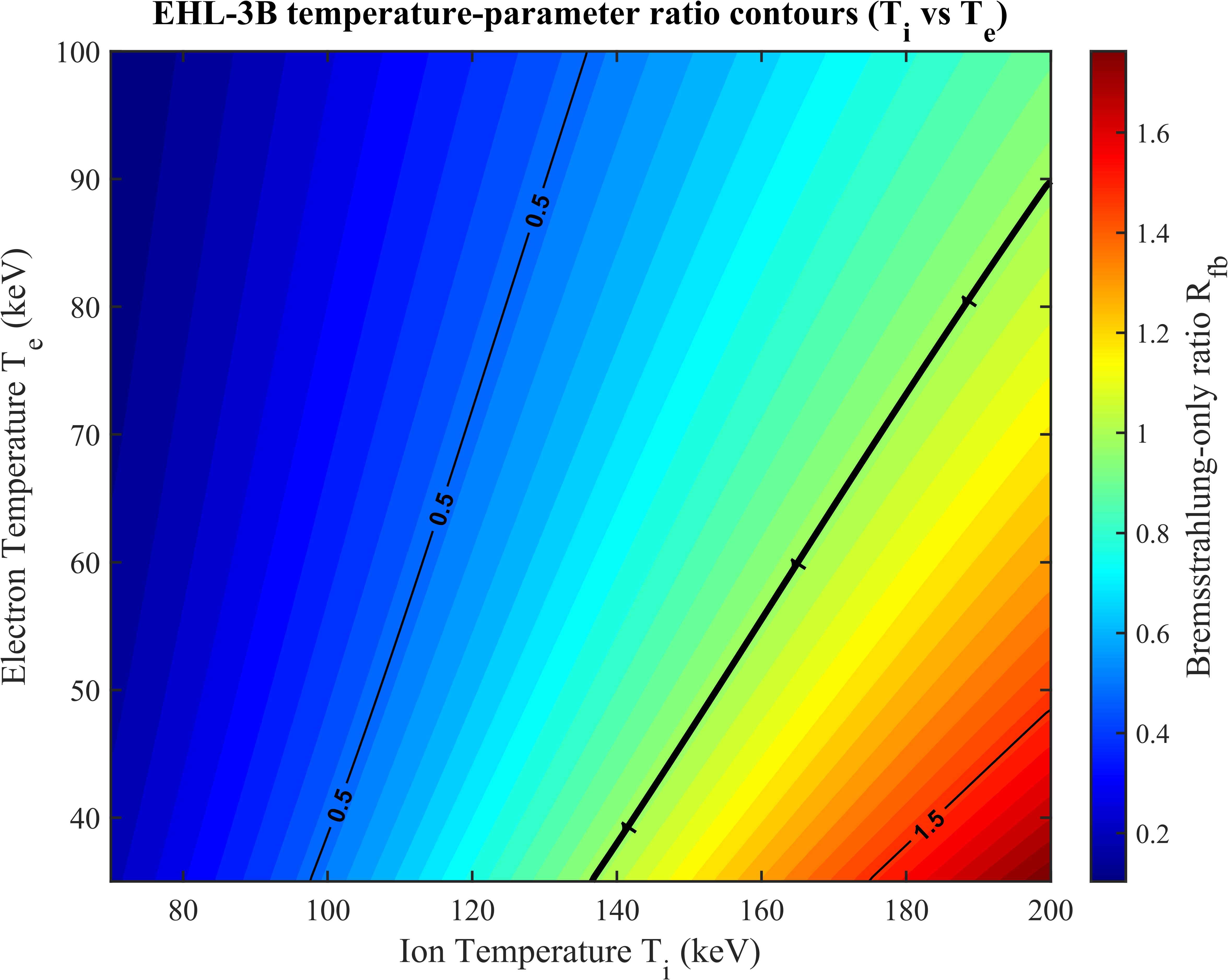}
        \caption{Drift-Maxwellian fusion-power evaluation}
    \end{subfigure}
    \caption{$\mathcal{R}_{\mathrm{fb}}$ as a function of $T_i$ and $T_e$ for EHL-3B at $\Omega_{p0}=0.90$~Mrad/s and $\gamma_{Bp}=0.20$. The drift-Maxwellian case gives the largest bremsstrahlung-only $\mathcal{R}_{\mathrm{fb}}$ values in the hot-ion, cooler-electron part of the prescribed-temperature parameter space.}
    \label{fig:popcon_temp}
\end{figure}

Additional scans over the fuel mixture ratio $n_B/n_p$ (Fig.~\ref{fig:popcon_fuel}) and the temperature pair $(T_i, T_e)$ (Fig.~\ref{fig:popcon_temp}) further delimit the model parameter region. Figure~\ref{fig:popcon_fuel}(a) shows the Maxwellian fusion-power evaluation, for which $\mathcal{R}_{\mathrm{fb}}$ remains below unity. Larger boron fractions increase $Z_{\mathrm{eff}}$ and therefore increase bremsstrahlung. Figure~\ref{fig:popcon_fuel}(b) uses the drift-Maxwellian reaction-rate coefficient in the fusion-power integral. In that case, the bremsstrahlung-only $\mathcal{R}_{\mathrm{fb}}>1$ region shifts toward $n_B/n_p \approx 0.080$--$0.15$ and sufficiently large $\Omega_{p0}$. The temperature scan in Fig.~\ref{fig:popcon_temp} indicates that the corresponding region is associated with hot ions and comparatively cooler electrons, near the baseline $T_i/T_e \approx 1.4\times10^2/35$~keV.

\section{Mechanism Analysis}
\label{sec:mechanisms}

The parameter scans in Section~\ref{sec:parameter} show two main trends. First, bremsstrahlung is reduced when relative toroidal flow is introduced. Second, at sufficiently large major radius, fusion power exhibits a non-monotonic dependence on $\gamma_{Bp}$ and can recover at large relative toroidal flow. The mechanisms responsible for these trends are discussed below.

\subsection{Reduction of Centrifugal Polarization}

\begin{figure}[!htbp]
    \centering
    \includegraphics[width=0.9\textwidth]{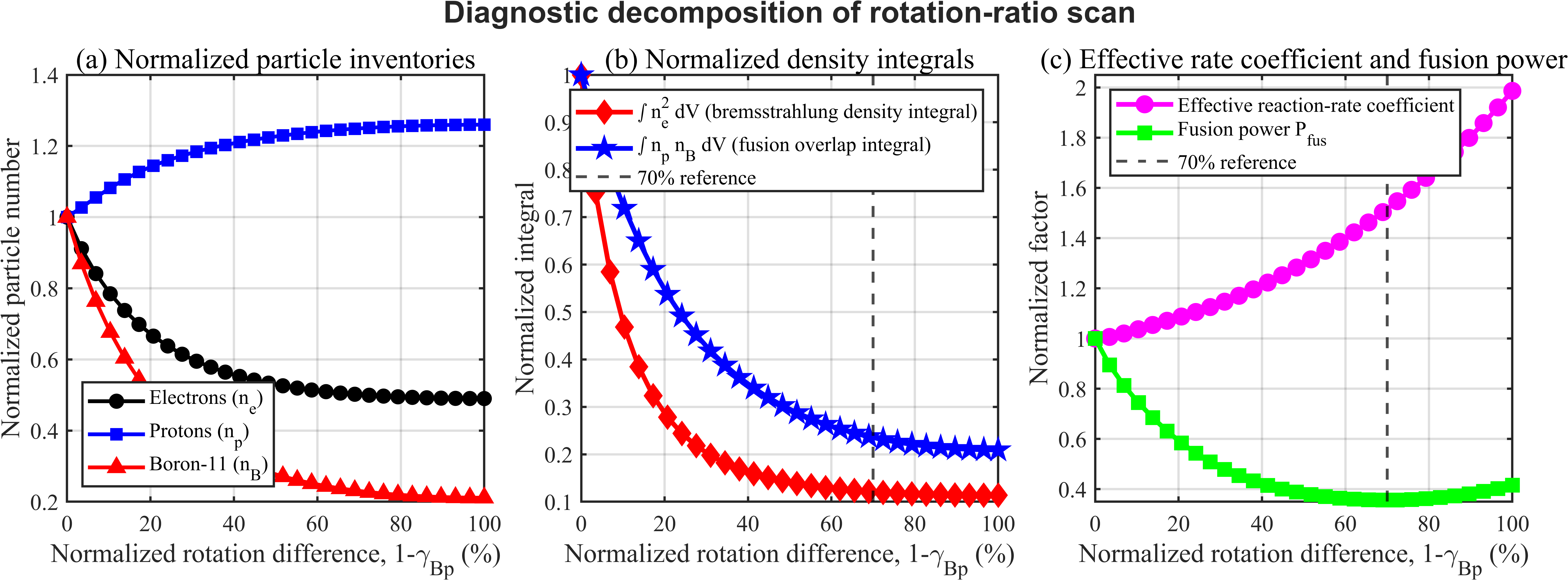}
    \caption{Diagnostic decomposition of contributing effects for the EHL-3B differential-rotation scan at fixed $\Omega_{p0}=0.90$~Mrad/s. (a) Normalized total particle inventories of electrons, protons, and boron. (b) Normalized geometric integral factors governing bremsstrahlung and fusion-density overlap. (c) Normalized density-overlap-weighted effective reaction-rate coefficient and normalized total fusion power. The proton--boron overlap integral decreases monotonically with increasing differential rotation, whereas the effective coefficient rises steadily; the partial recovery of $P_{\mathrm{fus}}$ at large differential rotation is associated with the competition between these two effects.}
    \label{fig:mechanism_inventory}
\end{figure}

Figure~\ref{fig:mechanism_inventory}(a) shows the global particle inventories $N_s=\int n_s\,dV$, normalized to their values at $\gamma_{Bp}=1$, for the EHL-3B species-dependent-rotation scan. As $\gamma_{Bp}$ decreases, the boron inventory $N_B$ drops because the smaller $\Omega_B$ weakens centrifugal accumulation on the low-field side (LFS). The electron inventory $N_e$ follows the same trend. Meanwhile, the proton inventory $N_p$ increases. With the core density fixed, a weaker polarization field $\Phi$ relaxes the electrostatic confinement of protons and allows the proton profile to broaden.

These changes affect the two power channels differently. Bremsstrahlung depends on $\int n_e^2\,dV$, whereas the fusion-density overlap is governed by $\int n_p n_B\,dV$. Figure~\ref{fig:mechanism_inventory}(b) shows that both integrals decrease as the differential rotation increases, but the decrease in $\int n_e^2\,dV$ is larger. This reflects the stronger sensitivity of the bremsstrahlung channel to the removal of high-density electrons from the LFS. By contrast, the proton--boron overlap integral decreases more gradually because the loss of boron is partly offset by the broadening of the proton distribution.

This difference explains why $P_{\mathrm{brem}}$ falls faster than $P_{\mathrm{fus}}$ as $\gamma_{Bp}$ is reduced from unity, and why $\mathcal{R}_{\mathrm{fb}}$ initially rises. At this stage, the increase in $\mathcal{R}_{\mathrm{fb}}$ is primarily a consequence of reduced centrifugal polarization and of the unequal response of the quadratic electron-density integral and the proton--boron overlap integral. The subsequent recovery of $P_{\mathrm{fus}}$ at larger relative toroidal flow requires an increase in the reaction-rate coefficient entering the fusion-power integral, which is analyzed in the next subsection.

\subsection{Drift Kinetic Energy and Fusion-Power Response}

For mechanism analysis, we define a density-overlap-weighted effective reaction-rate coefficient
\begin{equation}
\langle\sigma v\rangle_{\mathrm{eff}}
\equiv
\frac{P_{\mathrm{fus}}}
{E_{\mathrm{fus}}\int_V n_p n_B\, dV},
\label{eq:effective_rate_coefficient}
\end{equation}
which is obtained by dividing the volume-integrated fusion power by the corresponding proton--boron overlap integral and the energy released per reaction, $E_{\mathrm{fus}}$. This coefficient is the proton--boron-overlap-weighted average of the local $\langle\sigma v\rangle_{\mathrm{DM}}$ used in the post-processing. It is not a local reaction-rate coefficient and is not an independently evolved kinetic variable.

For the differential-rotation scan, Fig.~\ref{fig:mechanism_inventory}(c) plots the normalized effective reaction-rate coefficient
\begin{equation}
\widetilde{\langle\sigma v\rangle}_{\mathrm{eff}}(\gamma_{Bp})
=
\frac{\langle\sigma v\rangle_{\mathrm{eff}}(\gamma_{Bp})}
{\langle\sigma v\rangle_{\mathrm{eff}}(\gamma_{Bp}=1)},
\label{eq:norm_effective_rate_coefficient}
\end{equation}
and the normalized fusion power
\begin{equation}
\widetilde{P}_{\mathrm{fus}}(\gamma_{Bp})
=
\frac{P_{\mathrm{fus}}(\gamma_{Bp})}
{P_{\mathrm{fus}}(\gamma_{Bp}=1)}.
\label{eq:norm_fusion_power}
\end{equation}
These normalized quantities are used only to compare mechanisms across the scan and do not represent additional independently evolved state variables. The $\gamma_{Bp}=1$ case is a finite iso-rotation reference, so the normalization measures relative changes from the iso-rotation baseline rather than an absolute reaction-rate-coefficient scale.

\begin{figure}[!htbp]
    \centering
    \includegraphics[width=0.8\textwidth]{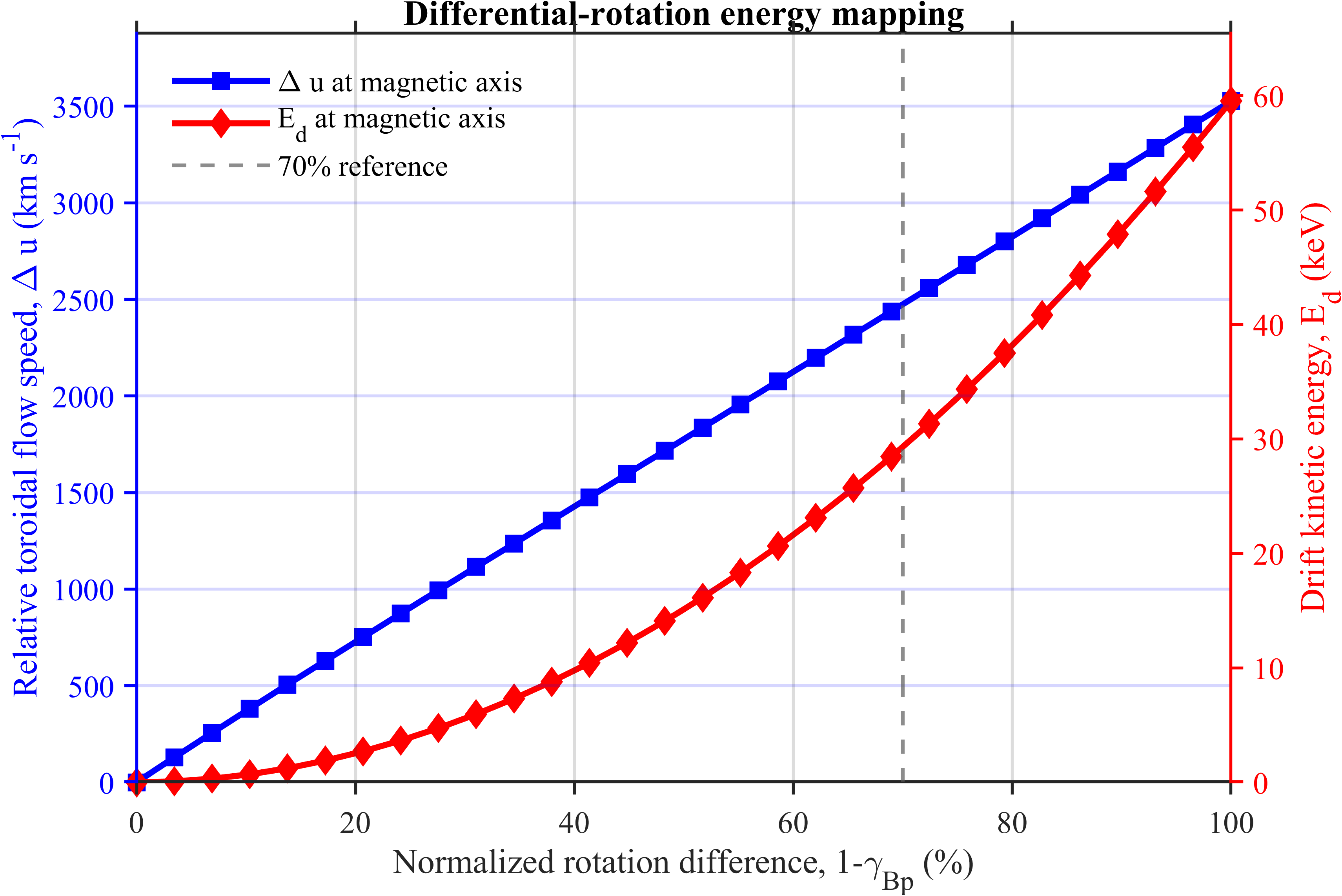}
    \caption{Relative toroidal flow speed $\Delta u$ and drift kinetic energy $E_{\mathrm{d}}$ at the magnetic axis versus the normalized rotation difference, $1-\gamma_{Bp}$, for EHL-3B at $\Omega_{p0}=0.90$~Mrad/s.}
    \label{fig:drift_energy_mapping}
\end{figure}

The relative toroidal flow speed $\Delta u = R |\Omega_p - \Omega_B|$ corresponds to a drift kinetic energy
\[
E_{\mathrm{d}} = \frac{1}{2} m_{\mathrm{r}} (\Delta u)^2
\]
in the center-of-mass frame. Figure~\ref{fig:drift_energy_mapping} shows $E_{\mathrm{d}}$ at the magnetic axis as a function of the normalized rotation difference, $1-\gamma_{Bp}$, for EHL-3B with $\Omega_{p0}=0.90$~Mrad/s. As $\gamma_{Bp}$ decreases, $1-\gamma_{Bp}$ increases and $E_{\mathrm{d}}$ rises quadratically, reaching about $60$~keV at $\gamma_{Bp}=0.10$.

\begin{figure}[!htbp]
    \centering
    \includegraphics[width=0.9\textwidth]{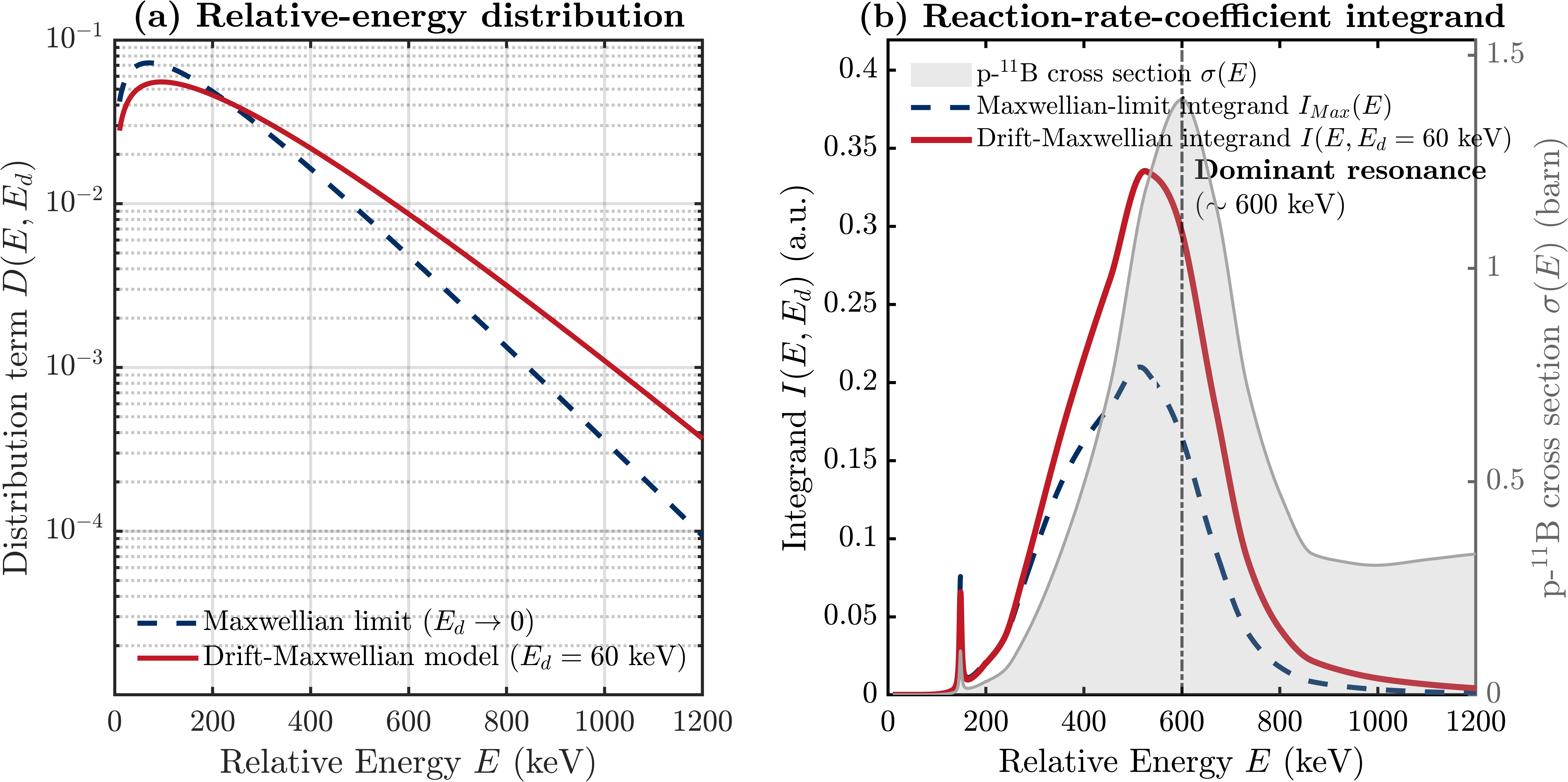}
    \caption{(a) Drift-Maxwellian distribution for $E_{\mathrm{d}}=0$ and $E_{\mathrm{d}}=60$~keV. (b) Integrand of the reaction-rate-coefficient integral, showing the finite-$E_{\mathrm{d}}$ shift toward the resonant energy range.}
    \label{fig:gamow_peak_excitation}
\end{figure}

Figure~\ref{fig:gamow_peak_excitation} illustrates the origin of the change in the reaction-rate coefficient. Panel (a) compares the drift-Maxwellian distribution $D_{\mathrm{DM}}(E,E_{\mathrm{d}})$ for the Maxwellian limit $E_{\mathrm{d}}=0$ and for $E_{\mathrm{d}}=60$~keV. The $\sinh$ factor in Eq.~\eqref{eq:drift_rate_coefficient} raises the high-energy tail when $E_{\mathrm{d}}$ is finite. Panel (b) shows the integrand of the reaction-rate-coefficient integral, $\sigma(E)\cdot D_{\mathrm{DM}}(E,E_{\mathrm{d}})\cdot \sqrt{E}$. In the purely thermal case, the integrand peaks near $1.0\times10^2$~keV and has little weight near the $6.0\times10^2$~keV resonance. When $E_{\mathrm{d}}=60$~keV, the high-energy tail increases and the integrand develops a secondary peak that overlaps the $p\text{-}^{11}\text{B}$ cross section. In this prescribed drift-Maxwellian model, $\langle \sigma v \rangle_{\mathrm{DM}}$ is therefore increased by a factor of several relative to the Maxwellian value at the same $T_i$.

The role of the reaction-rate coefficient can be seen more directly by combining Figs.~\ref{fig:mechanism_inventory}(b) and \ref{fig:mechanism_inventory}(c). Figure~\ref{fig:mechanism_inventory}(b) shows that the proton--boron overlap integral $\int n_p n_B\, dV$ decreases monotonically as the differential rotation increases, reflecting the continued reduction of centrifugal polarization. In contrast, Fig.~\ref{fig:mechanism_inventory}(c) shows that $\widetilde{\langle\sigma v\rangle}_{\mathrm{eff}}$ rises steadily with increasing differential rotation. The normalized total fusion power $\widetilde{P}_{\mathrm{fus}}$ therefore first decreases and then recovers.

The fusion power can therefore be written as the product of the density-overlap and reaction-rate-coefficient terms,
\[
P_{\mathrm{fus}} = E_{\mathrm{fus}}\left(\int n_p n_B\, dV\right)\langle\sigma v\rangle_{\mathrm{eff}}.
\]
At moderate relative toroidal flow, the decrease in the proton--boron overlap integral dominates and $P_{\mathrm{fus}}$ falls. At larger relative toroidal flow, the rise of $\langle\sigma v\rangle_{\mathrm{eff}}$ offsets part of the continuing density penalty, and $P_{\mathrm{fus}}$ correspondingly recovers from its minimum. The recovery of $P_{\mathrm{fus}}$ at small $\gamma_{Bp}$ thus cannot be attributed to a larger proton--boron overlap; it reflects the competition between decreasing fusion-density overlap and an increasing effective reaction-rate coefficient.

This interpretation is consistent with the drift-Maxwellian model shown in Fig.~\ref{fig:gamow_peak_excitation}: the increasing $E_{\mathrm{d}}$ shifts weight toward the resonant part of the $p\text{-}^{11}\text{B}$ cross section, thereby increasing the reaction-rate coefficient without requiring a corresponding increase in the prescribed thermal temperature profile.

\subsection{Scale Dependence}

Within the accessible rotation range of each device, the difference between EHL-2 and EHL-3B is consistent with the larger $R^2$ drift-energy scale in the larger geometry. The two configurations also differ in baseline density, temperature, magnetic field, and plasma current, so this comparison should not be read as a controlled single-variable radius scan. For a given $\Delta\Omega$, one has $E_{\mathrm{d}} \propto R^2$. In EHL-2 ($R_0 = 1.1$~m), $E_{\mathrm{d}}$ remains below about $5.0$~keV even for the largest relative toroidal flows considered, which is too small to modify the drift-Maxwellian reaction-rate coefficient appreciably. The increase in $\mathcal{R}_{\mathrm{fb}}$ comes mainly from the reduction in bremsstrahlung.

In EHL-3B ($R_0 = 3.2$~m), the larger radius increases $E_{\mathrm{d}}$ by nearly an order of magnitude, so the reaction-rate-coefficient contribution to fusion power becomes visible in the scan. The proton--boron overlap integral still decreases with increasing relative toroidal flow, but the effective coefficient grows enough for $P_{\mathrm{fus}}$ to recover from its intermediate-$\gamma_{Bp}$ minimum. Because $P_{\mathrm{brem}}$ decreases much more strongly, the bremsstrahlung-only $\mathcal{R}_{\mathrm{fb}}$ can reach the $\mathcal{R}_{\mathrm{fb}}=1$ level within the present model parameter set.

\section{Conclusions}
\label{sec:conclusion}

This work uses rotating $p\text{-}^{11}\text{B}$ ST plasmas as a demanding model case for parameter-scan VEQ-MF multi-fluid equilibrium calculations. Under prescribed species-rotation and temperature profiles, the solver was benchmarked against a finite-difference reference and then used to map the $(\Omega_p,\Omega_B)$ parameter space for the EHL-2 and EHL-3B geometries. Three in-range FDM/VEQ-MF benchmark points give few-percent agreement in the global quantities used for the power-ratio analysis, and a representative in-range VEQ-MF convergence check shows sub-percent sensitivity to modal and quadrature refinement. The physical discussion therefore represents model behavior within this reduced equilibrium-and-post-processing calculation, rather than a reactor-feasibility assessment.

Under iso-rotation ($\Omega_p=\Omega_B$), the large mass disparity between protons and boron drives a pronounced low-field-side accumulation of boron. The resulting electrostatic polarization shifts electrons outward, increases $\int n_e^2\,dV$, and causes bremsstrahlung to rise faster than fusion power. Within the present model, this leads to a systematic reduction of $\mathcal{R}_{\mathrm{fb}}$ with increasing rotation in both geometries.

Species-dependent toroidal rotation ($\Omega_p \neq \Omega_B$) changes this balance through two distinct contributions. First, reducing $\Omega_B$ relative to $\Omega_p$ weakens the centrifugal forcing on boron, reduces the poloidal density asymmetry, and lowers the electron-density peaking on the low-field side. This decreases the bremsstrahlung-driving integral $\int n_e^2\,dV$ and therefore reduces $P_{\mathrm{brem}}$. Second, the same relative toroidal flow generates a finite drift kinetic energy $E_{\mathrm{d}}$, which increases the adopted drift-Maxwellian reaction-rate coefficient. The diagnostic decomposition shows that the proton--boron overlap integral $\int n_p n_B\,dV$ decreases monotonically as the differential rotation increases, whereas the density-overlap-weighted effective coefficient rises steadily. The recovery of $P_{\mathrm{fus}}$ at large differential rotation is therefore associated with the competition between decreasing fusion-density overlap and an increasing reaction-rate-coefficient contribution to fusion power.

This competition is weak in EHL-2 because the smaller major radius limits the attainable drift kinetic energy. As a result, the increase in $\mathcal{R}_{\mathrm{fb}}$ there is associated mainly with the reduction in bremsstrahlung. In EHL-3B, the larger major radius gives a larger drift-energy scale, making the reaction-rate-coefficient contribution to fusion power visible within the chosen baseline. Within the present reduced model and prescribed-profile assumptions, the fixed-$\Omega_{p0}=0.90$~Mrad/s rotation-ratio scan raises the bremsstrahlung-only $\mathcal{R}_{\mathrm{fb}}$ from 0.18 at $\gamma_{Bp}=1$ to about 1.3 at $\gamma_{Bp}=0$. The fixed-$\gamma_{Bp}=0.20$ rotation-strength scan reaches the same bremsstrahlung-only crossing level near the upper end of the reported scan range. The two-dimensional maps place the $\mathcal{R}_{\mathrm{fb}}=1$ contour in the high-$\Omega_{p0}$, low-$\gamma_{Bp}$ part of the scanned parameter space, with fuel-mixture and temperature scans favoring intermediate boron fraction and hot ions with cooler electrons.

Several modeling limitations should be kept in mind when interpreting these results. Poloidal flow inertia is neglected even though the highest-rotation cases enter the transonic or supersonic boron range. The scalar temperature profiles are prescribed as flux functions, and the rotation profiles are imposed rather than evolved from torque balance. Inter-species collisional momentum exchange, frictional thermalization, relative-flow-driven microinstabilities, and transport-level profile evolution are not included. The scan also covers only the $\Omega_p\geq\Omega_B$ branch, and its upper boundary is set by numerical convergence of the prescribed-flow GS solve rather than by a physical stability criterion.

These cases correspond to $\mathcal{R}_{\mathrm{fb}}=1$ crossings of a reduced bremsstrahlung-only power ratio obtained under prescribed rotation and temperature profiles. Other radiative and transport losses, inter-species momentum exchange, frictional thermalization, and auxiliary power are not included in that ratio. For the largest imposed relative-toroidal-flow EHL-3B case, the collisional momentum-exchange power scale is about $5.2$~GW, whereas the post-processed fusion power is about $1.2\times10^2$~MW. This momentum-exchange estimate is a scope check consistent with Rider-type sustainment concerns, not a separate feasibility argument. It reinforces that the main contribution is a benchmarked prescribed-profile scan calculation and diagnostic analysis of contributing effects, not a claim of route feasibility. A natural next step is to couple the equilibrium calculation with transport-level analyses of robustness, temperature evolution, and the energetic cost of sustaining species-dependent toroidal rotation.

\appendix
\titleformat{\section}
  {\normalfont\Large\bfseries}
  {Appendix \thesection:}
  {0.5em}
  {}

\section{Prescribed Profiles}
\label{app:profile_sensitivity}

The multi-fluid equilibrium model requires reference scalar profiles as functions of the normalized poloidal flux $\bar{\psi} = (\psi - \psi_0)/(\psi_a - \psi_0)$. The following power-polynomial forms are used. Their baseline exponents are chosen to isolate the effect of centrifugal multi-fluid polarization:
\begin{align}
    N_s(\bar{\psi}) &= N_{s0} \left[ \epsilon_n + (1-\epsilon_n)(1 - \bar{\psi}^{\beta_n})^{\alpha_n} \right], \label{eq:prof_n} \\
    T_s(\bar{\psi}) &= T_{s0} \left[ \epsilon_T + (1-\epsilon_T)(1 - \bar{\psi}^{\beta_T})^{\alpha_T} \right], \label{eq:prof_t} \\
    \Omega_s(\bar{\psi}) &= \Omega_{s0} (1 - \bar{\psi}^{\beta_\Omega})^{\alpha_\Omega}, \label{eq:prof_om} \\
    F(\bar{\psi}) &= R_0 B_0 \left[ 1 - \gamma_f (1 - \bar{\psi}^{\beta_f})^{\alpha_f} \right].
    \label{eq:prof_f}
\end{align}
Here $N_{s0}$, $T_{s0}$, and $\Omega_{s0}$ are the on-axis reference density, temperature, and rotation frequency of species $s$. The parameters $\epsilon_n$, $\epsilon_T$, $\alpha_i$, $\beta_i$, and $\gamma_f$ control the profile shape.

\subsection{Temperature and Density Profiles}

\begin{figure}[!htbp]
    \centering
    \includegraphics[width=0.9\textwidth]{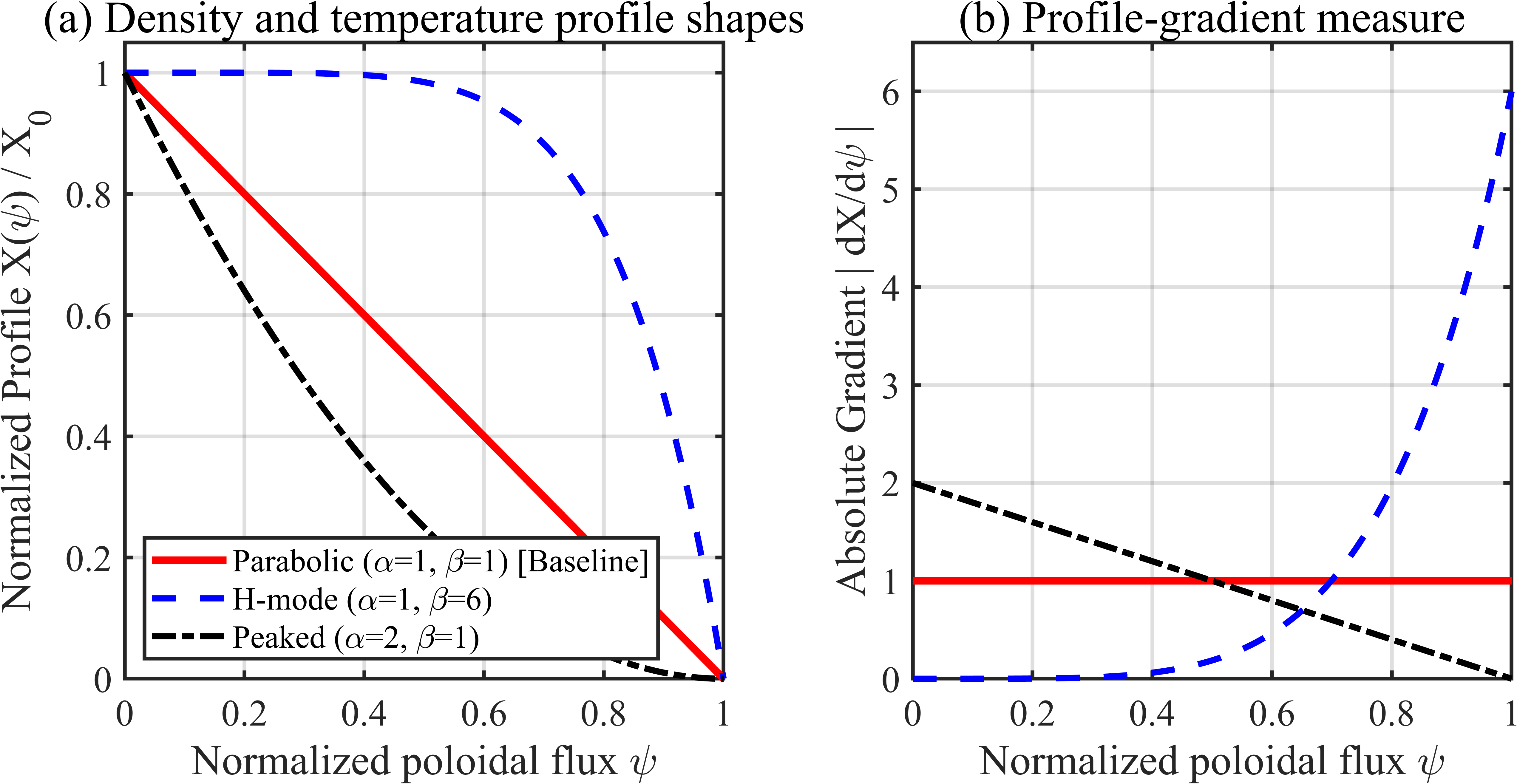}
    \caption{Comparison of temperature and density profiles (left) and the corresponding radial-gradient penalties (right). The parabolic baseline, shown in red, retains a broad fusion volume while avoiding a sharp edge gradient.}
    \label{fig:app_nt_profiles}
\end{figure}

The baseline temperature and density profiles are taken to be parabolic, with $\alpha_{n,T} = 1.0$ and $\beta_{n,T} = 1.0$, so that $X(\bar{\psi}) \propto (1-\bar{\psi})$. Figure~\ref{fig:app_nt_profiles} compares this baseline with more strongly peaked and step-like alternatives. The baseline preserves an adequate fusion volume while avoiding overly large pressure gradients in strongly rotating cases.

A strongly peaked profile confines the highest-temperature region to a narrow core and reduces the reacting volume. A step-like H-mode profile, on the other hand, introduces a large edge gradient that can affect the ellipticity of the GS equation at high rotation. The parabolic baseline avoids a sharp edge structure while retaining a sufficiently broad reacting volume for the fusion-power integral.

\subsection{Rotation Profile}

\begin{figure}[!htbp]
    \centering
    \includegraphics[width=0.9\textwidth]{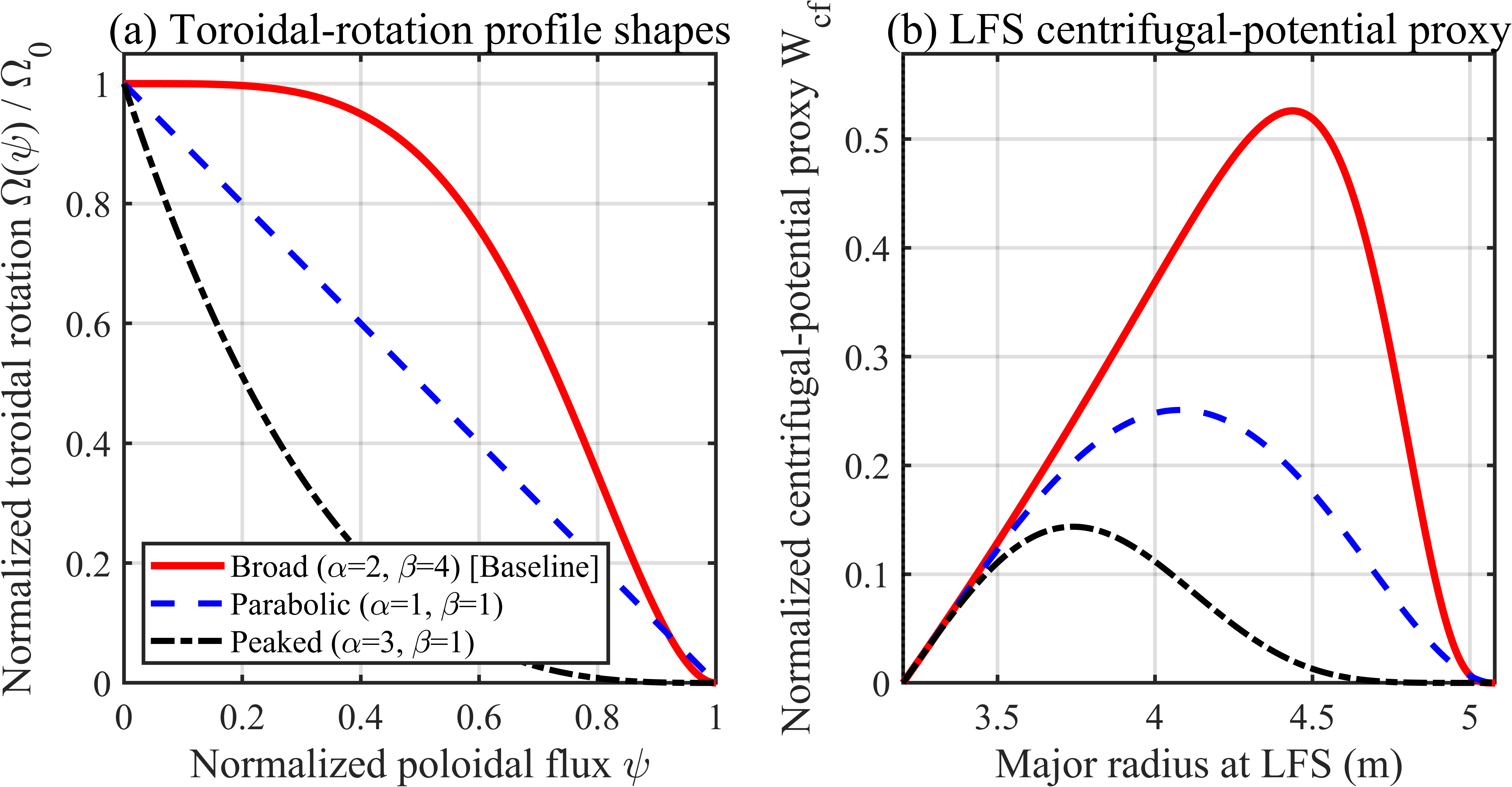}
    \caption{Sensitivity of centrifugal polarization to variations of the rotation profile: (a) normalized rotation profiles and (b) LFS centrifugal potential $W_{\mathrm{cf}}$.}
    \label{fig:app_rot_profiles}
\end{figure}

The angular-velocity profile $\Omega_s(\bar{\psi})$ controls the strength of centrifugal species redistribution. In this work, a broad, nearly rigid-body profile is used, with $\alpha_\Omega = 2.0$ and $\beta_\Omega = 4.0$, as an idealized profile motivated by neutral-beam-injection-like momentum deposition. Figure~\ref{fig:app_rot_profiles} compares the baseline with narrower rotation profiles and shows the corresponding change in the LFS centrifugal potential.

If the profile is too strongly peaked, the rotation decays before reaching the LFS, which reduces both the radiation penalty and the spatial extent over which drift kinetic energy is available. The broader baseline profile keeps the rotation appreciable in the large-radius region and provides a clearer comparison between iso-rotation and differential-toroidal-rotation cases.

\section{High-Rotation Equilibrium Limit}
\label{app:rotation_limit}

For EHL-3B, the central angular frequency is limited to $\Omega_{\mathrm{max}} = 0.90 \times 10^6$~rad/s. Above this value, the equilibrium solution no longer remains well behaved under the strong centrifugal loading considered here.

\subsection{Centrifugal Potential Amplification}

\begin{figure}[!htbp]
    \centering
    \includegraphics[width=0.8\textwidth]{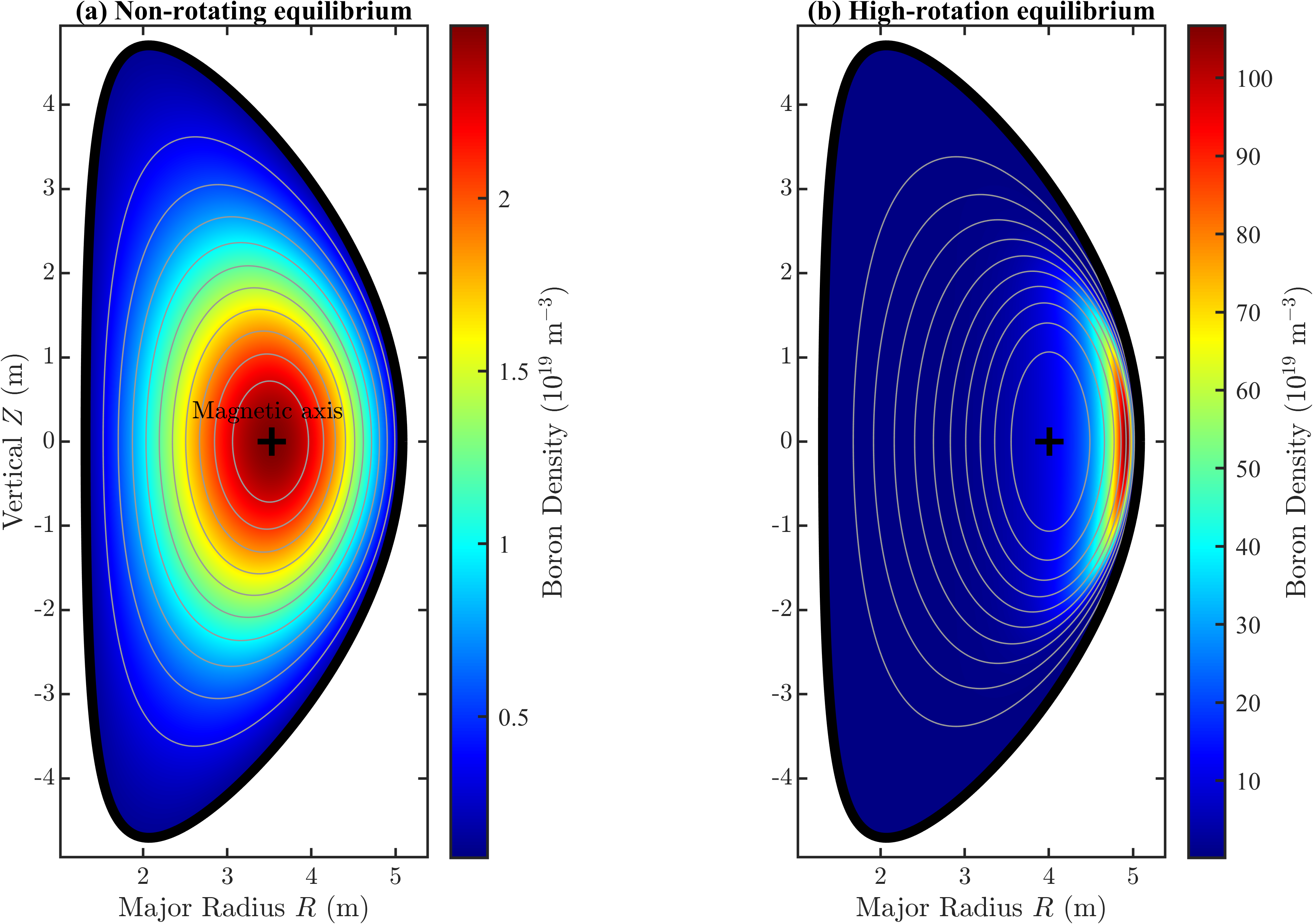}
    \caption{Boron density $n_B$: (a) static equilibrium and (b) near the critical rotation limit, $\Omega = 1.0$~Mrad/s.}
    \label{fig:app_B_density}
\end{figure}

In EHL-3B, the centrifugal potential
\[
W_s^{\mathrm{cf}} = \frac{1}{2} m_s \Omega_s(\bar{\psi})^2 (R^2 - R_0^2)
\]
becomes large on the LFS. For $^{11}\text{B}$ approaching $1.0$~Mrad/s, the heavy-ion population enters a strongly rotating regime. The HFS is depleted of boron, while boron accumulates in a narrow layer near the LFS boundary, as shown in Fig.~\ref{fig:app_B_density}.

\subsection{Shafranov Shift and Flux Compression}

\begin{figure}[!htbp]
    \centering
    \includegraphics[width=0.8\textwidth]{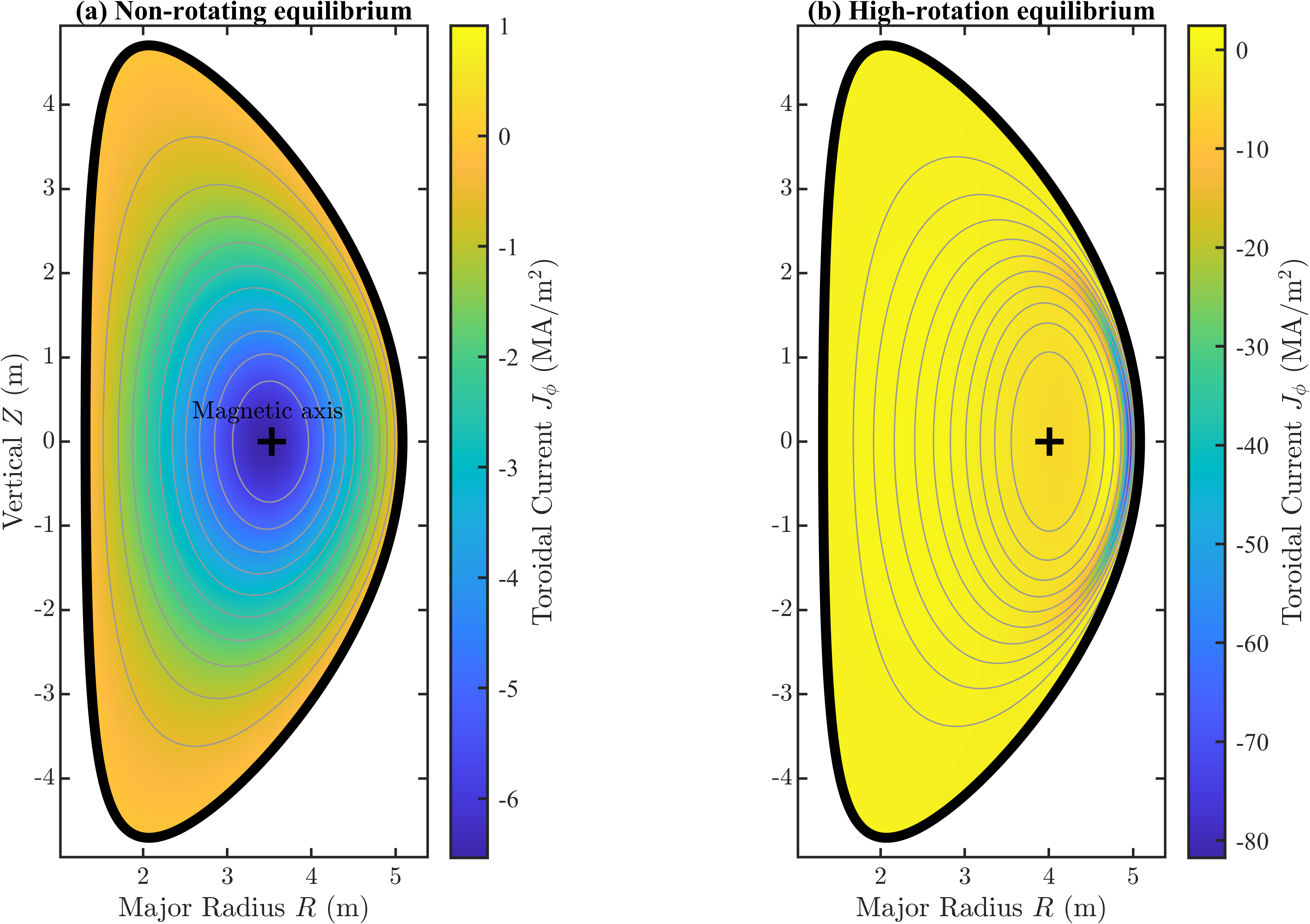}
    \caption{Toroidal current density $J_\phi$ and flux-surface topology: (a) static equilibrium and (b) $\Omega = 1.0$~Mrad/s.}
    \label{fig:app_B_current}
\end{figure}

Under strong centrifugal inertia, the thermal and dynamic pressures become concentrated on the LFS, which drives a large outward Shafranov shift. Because the last closed flux surface cannot expand indefinitely within the fixed boundary, the flux surfaces become compressed on the outboard side, as shown in Fig.~\ref{fig:app_B_current}.

\subsection{Numerical Solver Response}

Beyond $\Omega_{\mathrm{max}}$, the VEQ-MF iterations either diverge or converge to nonphysical states because the GS operator loses elliptic robustness under the imposed centrifugal loading. For this reason, the evaluation is restricted to $0.90$~Mrad/s in order to retain a physically meaningful flux-surface structure.

\section{Momentum-Exchange Power Estimate}
\label{app:momentum_exchange}

The parameter scans impose $\Omega_p(\psi)$ and $\Omega_B(\psi)$ independently. In a collisional plasma, maintaining finite relative toroidal flow between protons and boron would require external torque against inter-species momentum exchange. To estimate the missing power scale, we use the linearized Landau--Spitzer unlike-ion collision frequency in SI units~\cite{huba2019nrl},
\begin{equation}
    \nu_{pB} =
    \frac{4\sqrt{2\pi}}{3}
    \frac{n_B Z_p^2 Z_B^2 e^4 \ln\Lambda}
    {(4\pi\epsilon_0)^2 m_p m_{\mathrm{r}}}
    \left(
    \frac{k_B T_p}{m_p} + \frac{k_B T_B}{m_B}
    \right)^{-3/2},
    \label{eq:nu_pB_estimate}
\end{equation}
where $m_{\mathrm{r}}=m_p m_B/(m_p+m_B)$, $Z_p=1$, and $Z_B=5$. The corresponding frictional power density required to hold the imposed relative flow is estimated as
\begin{equation}
    p_{\mathrm{fric}} \simeq m_p n_p \nu_{pB}
    \left[ R(\Omega_p-\Omega_B) \right]^2,
    \qquad
    P_{\mathrm{fric}}=\int_V p_{\mathrm{fric}}\,dV .
    \label{eq:pfric_estimate}
\end{equation}
This expression is used only as an order-of-magnitude estimate; it is not a self-consistent transport calculation and is not added as a corrected power-balance denominator. The dissipated directed kinetic energy would enter the ion energy balance and modify the temperature profiles, which are prescribed in the present model.

Table~\ref{tab:momentum_exchange_estimate} evaluates Eqs.~\eqref{eq:nu_pB_estimate}--\eqref{eq:pfric_estimate} for the largest imposed relative-toroidal-flow EHL-3B point in Fig.~\ref{fig:hl3b_gamma_scan}, using $\ln\Lambda=10$ and the same two-dimensional equilibrium fields used for the power integrals. The estimate gives $P_{\mathrm{fric}}\simeq 5.2$~GW, about $44$ times the post-processed fusion power at the same point. This indicates that the $\mathcal{R}_{\mathrm{fb}}=1$ result of the bremsstrahlung-only power ratio is a prescribed-profile mechanism result rather than a self-consistent sustainment estimate.

\begin{table}[!htbp]
    \centering
    \caption{Order-of-magnitude proton--boron momentum-exchange estimate for the largest imposed relative-toroidal-flow EHL-3B case.}
    \label{tab:momentum_exchange_estimate}
    \begin{tabular}{lc}
        \toprule
        \textbf{Quantity} & \textbf{Estimate} \\
        \midrule
        Case & $\Omega_{p0}=0.90$~Mrad/s, $\Omega_{B0}=0$ \\
        Volume-averaged densities & $\langle n_p\rangle_V=9.2\times10^{19}$~m$^{-3}$, $\langle n_B\rangle_V=8.6\times10^{18}$~m$^{-3}$ \\
        Maximum relative speed & $\max[R|\Omega_p-\Omega_B|]=4.0\times10^6$~m/s \\
        Mass-weighted collision frequency & $\bar{\nu}_{pB}\simeq 1.1\times10^1$~s$^{-1}$ \\
        Frictional sustainment power & $P_{\mathrm{fric}}\simeq 5.2\times10^3$~MW \\
        Post-processed fusion power & $P_{\mathrm{fus}}\simeq 1.2\times10^2$~MW \\
        Ratio & $P_{\mathrm{fric}}/P_{\mathrm{fus}}\simeq 44$ \\
        \bottomrule
    \end{tabular}
\end{table}

\section{VEQ-MF Numerical Methods}
\label{app:numerical_solver}

This appendix summarizes the numerical treatments in VEQ-MF that allow stable convergence under strong centrifugal loading.

\subsection{Shifted Chebyshev Expansion}

VEQ-MF maps the physical domain $(R,Z)$ to normalized coordinates $(\rho,\theta)$ and expands the geometric shape functions and $\psi$ in shifted Chebyshev polynomials:
\begin{equation}
    f(\rho) = \rho^p \left[ f_a + \sum_{l=0}^{L} f_l (1-\rho^2) T_l(2\rho^2-1) \right],
    \label{eq:chebyshev_expansion_app}
\end{equation}
where $T_l$ is the Chebyshev polynomial of the first kind. The GS residuals are then projected by matrix-kernel acceleration,
\begin{equation}
    \mathcal{R} = M^T \cdot (\mathbf{W} \odot \mathbf{G}_{grid}),
    \label{eq:matrix_kernel_app}
\end{equation}
where $\mathbf{G}_{grid}$ is the physical error vector on a Gaussian quadrature grid, $M^T$ is the precomputed projection matrix, and $\mathbf{W}$ contains the quadrature weights.

\subsection{Hybrid Nonlinear Solver with SVD Regularization}

The nonlinear solve uses a rank-one Broyden update together with a backtracking line search. When the line search fails, an exact finite-difference Jacobian is constructed and inverted with an SVD pseudo-inverse. The pseudo-inverse retains singular values $s_i$ satisfying $s_i>\max(10^{-6}s_{\max},10^{-9})$ and sets the reciprocals of smaller singular values to zero. This regularization suppresses nonphysical spectral noise.

\section{Supplementary Modal Sensitivity Check}
\label{app:modal_sensitivity}

A supplementary modal-sensitivity check was carried out for the VEQ-MF stress-test input ($\Omega_{p0}=1.0\times10^6$~rad/s, $\Omega_{B0}=0.20\times10^6$~rad/s, $\gamma_{Bp}=0.20$, EHL-3B). Table~\ref{tab:modal_sensitivity} compares the standard order-3 Chebyshev representation with 16 spectral parameters against an order-5 representation with 24 spectral parameters, keeping the same prescribed profiles, rotation frequencies, 16$\times$16 Gaussian quadrature grid, and 25~MA current target. This point lies slightly outside the nominal scan boundary of $0.90$~Mrad/s and is reported only as a stress-test sensitivity check; the primary in-range convergence evidence is given in Table~\ref{tab:convergence_in_range}.

\begin{table}[!htbp]
    \centering
    \caption{Supplementary VEQ-MF modal-sensitivity check for the EHL-3B stress-test input. Absolute values are rounded for scale; relative differences are computed from the unrounded order-3 and order-5 runs.}
    \label{tab:modal_sensitivity}
    \begin{tabular}{lccccc}
        \toprule
        \textbf{Modal truncation} & \textbf{Parameters} & $W_{\mathrm{mhd}}$ (MJ) & $P_{\mathrm{brem}}$ (MW) & $P_{\mathrm{fus}}$ (MW) & $\mathcal{R}_{\mathrm{fb}}$ \\
        \midrule
        Order 3 & 16 & $1.7\times10^3$ & $1.0\times10^2$ & $1.2\times10^2$ & $1.1$ \\
        Order 5 & 24 & $1.7\times10^3$ & $1.0\times10^2$ & $1.2\times10^2$ & $1.1$ \\
        Relative difference & -- & 0.079\% & 0.11\% & 0.18\% & 0.064\% \\
        \bottomrule
    \end{tabular}
\end{table}

\section*{Acknowledgments}
The authors would like to thank Ruohan Zhang, Caixue Chen, Huibin Zhou, and Xinyu Liu for helpful discussions and kind assistance during this work. This research is supported by the National MCF Energy R\&D Program of China (Grant Nos. 2025YFF0512002 and 2022YFE03090000), the National Natural Science Foundation of China (Grant Nos. 12435014 and 12475214), the Fundamental Research Funds for the Central Universities of Ministry of Education of China (Grant No. DUT25Z2536), and the Natural Science Foundation of Liaoning (Grant No. 2025-MSLH-157).

\end{document}